\documentclass[preprint]{aastex}
\usepackage{epsf}
\usepackage{emulateapj5}

\slugcomment{Submitted to The Astrophysical Journal}

\newcommand\avg[1]{\langle{#1}\rangle}
\newcommand\mm{\mu_{\rm core}}
\newcommand\ma{\avg{\mm}}
\newcommand\E[1]{\times10^{#1}}
\newcommand\uu{{\bf u}}
\newcommand\xx{{\bf x}}

\newcommand\refeq[1]{eq.~(\ref{eq:#1})}
\newcommand\refeqs[2]{eqs.~(\ref{eq:#1}) and (\ref{eq:#2})}
\newcommand\refEqs[2]{Eqs.~(\ref{eq:#1}) and (\ref{eq:#2})}
\newcommand\reftab[1]{Table~\ref{tab:#1}}
\newcommand\reffig[1]{Fig.~\ref{fig:#1}}

\begin{document}

\twocolumn[
\title{Lensing and the Centers of Distant Early-Type Galaxies}
\author{Charles R. Keeton\altaffilmark{1}}
\affil{Astronomy and Astrophysics Department, University of Chicago,
  5640 S.\ Ellis Ave., Chicago, IL 60637}

\begin{abstract}
Gravitational lensing provides a unique probe of the inner 10--1000
pc of distant galaxies ($z \sim 0.2$--1). Lens theory predicts that
every strong lens system should have a faint image near the center
of the lens galaxy, which should be visible in radio lenses but
have not been observed. We study these ``core'' images using models
derived from the stellar distributions in nearby early-type galaxies.
We find that realistic galaxies predict a remarkably wide range of
core images, with lensing magnifications spanning some six orders
of magnitude. More concentrated galaxies produce fainter core
images, although not with any simple, quantitative, model independent
relation.  Some real galaxies have diffuse cores and predict bright
core images (magnification $\mm \gtrsim 0.1$), but more common are
galaxies that predict faint core images ($\mm \lesssim 0.001$).
Thus, stellar mass distributions alone are probably concentrated
enough to explain the lack of observed core images, and may require
observational sensitivity to improve by an order of magnitude before
detections of core images become common. Two-image lenses will tend
to have brighter core images than four-image lenses, so they will
be the better targets for finding core images and exploiting these
tools for studying the central mass distributions of distant galaxies.
\end{abstract}
]
\altaffiltext{1}{Hubble Fellow}

\section{Introduction}

Galaxy centers are interesting places to study dynamics and galaxy
formation. Their short crossing times make them sensitive to
dynamical processes such as relaxation and binary black hole
heating (e.g., Ebisuzaki, Makino \& Okumura 1991; Milosavljevic \&
Merritt 2001). Their deep potential wells collect remnants of the
galaxy formation process such as the cores of accreted galaxies
(e.g., de Zeeuw \& Franx 1991; Barnes \& Hernquist 1992). Their
dark matter content provides clues to the interaction between
baryons and dark matter by adiabatic compression during galaxy
formation (e.g., Blumenthal et al.\ 1986), and may even reveal
properties of the dark matter particle such as cross sections for
self-interactions (e.g., Spergel \& Steinhardt 2000).

Nearby, galaxy centers can be studied directly with high spatial
resolution observations. Hubble Space Telescope imaging of
early-type galaxies shows that, contrary to theoretical
expectations (e.g., Tremaine 1997), the luminosity profiles diverge
at small radii (e.g., Faber et al.\ 1997; Ravindranath et al.\ 2001;
Rest et al.\ 2001). The profiles seem to fall into two classes:
``core'' galaxies have a distinct transition between a steep outer
profile and a shallow inner core that has $I \propto R^{-\gamma}$
with $\gamma \lesssim 0.3$; while ``power law'' galaxies show no
such break and have steep central cusps with $\gamma \gtrsim 0.5$.
Interestingly, the global properties of the galaxies seem to
correlate well with the centers. Core galaxies tend to be luminous,
slowly rotating systems with boxy or elliptical isophotes, while
power law galaxies tend to be faint, rapidly rotating systems with
disky isophotes. Although the division may not be as sharp as
originally thought (see Ravindranath et al.\ 2001; Rest et al.\
2001), it still puts strong constraints on the formation process.
In hierarchical merging scenarios, simple models cannot easily
explain why the large galaxies are so much less dense than their
small progenitors (see \S7 of Faber et al.\ 1997, and references
therein), and some additional process such as heating by binary
black holes may be required (e.g., Milosavljevic \& Merritt 2001;
Milosavljevic et al.\ 2001).

For distant galaxies, we cannot directly resolve 10--100 parsec
scales, but we can instead turn to a unique indirect probe offered
by gravitational lensing. Lens theory predicts that if the central
mass distribution is shallower than $\rho \propto r^{-2}$ then any
multiply-imaged gravitational lens must have an odd number of images
(Burke 1981; Schneider, Ehlers \& Falco 1992). Standard image
configurations\footnote{The rare exception is a configuration where
the source lies in a naked cusp (e.g., Schneider et al.\ 1992); among
more than 60 known lenses, APM~08279+5255 is the only candidate naked
cusp lens (Lewis et al.\ 2002). The only other exception is B1359+154,
a unique lens where three lens galaxies jointly produce six bright
images, and models predict three additional core images (Rusin et
al.\ 2001).} have two or four bright images lying $\sim$3--10 kpc
from the center of the lens galaxy, with the remaining image just
10--100 pc from the center and much fainter than the others. Because
the ``core'' image is very sensitive to the central surface density
of the lens galaxy, with a higher density corresponding to a fainter
image, it offers a unique way to constrain the density on scales that
cannot be directly resolved. This probe of galaxy centers can in
principle be applied to all lens galaxies, which now number more than
60 and are predominantly early-type galaxies spanning the redshift
range $z \sim 0.3$--1 (e.g., Kochanek et al.\ 2000). It is conceptually
equivalent to using radial arcs to constrain the central profiles of
lensing clusters (e.g., Mellier, Fort \& Kneib 1993; Smail et al.\
1996; Molikawa \& Hattori 2001; Oguri, Taruya \& Suto 2001).

The best observational data on core images come from radio lenses,
because the lack of radio emission from most lens galaxies enables
sensitive searches for core images. The Cosmic Lens All-Sky Survey
found 18 radio lenses but no core images, based on radio maps where
the dynamic range is typically several tens to several hundreds but
reaches 1200 for B1030+074 and 2000 for B0218+357 (Rusin \& Ma 2001;
Norbury et al.\ 2002). Several other radio lenses have candidate core
images (MG~1131+0456, Chen \& Hewitt 1993; PMN J1632-0033, Winn et
al.\ 2002) although the hypothesis that the central radio flux
originates in the lens galaxy cannot be ruled out. At optical and
near-infrared wavelengths, APM~08279+5255 has an odd number of images
(Ibata et al.\ 1999; Egami et al.\ 2000), but its interpretation is
not clear. The third image may be a core image, in which case it
indicates a large low-density core in the lens galaxy (Ibata et al.\
1999; Egami et al.\ 2000; Mu\~noz, Kochanek \& Keeton 2001), or it
may be a case of a ``naked cusp'' image configuration, in which case
it contains no information about the center of the lens galaxy (Lewis
et al.\ 2002). No other optical core images have been seen, although
the searches are of course hindered by light from the lens galaxies.

The apparent discrepancy between data and theory provides the
desired opportunity to learn about the centers of distant galaxies.
Motivated both by theoretical expectations (see Tremaine 1997) and by
ease of use, many analyses have assumed models with a finite density
core and obtained limits on lens galaxy core radii
(e.g., Narayan, Blandford \& Natyananda 1984;
Narasimha, Subramanian \& Chitre 1986;
Blandford \& Kochanek 1987;
Hinshaw \& Krauss 1987;
Narayan \& Schneider 1990;
Wallington \& Narayan 1993;
Kochanek 1996;
Evans \& Hunter 2002).
Rusin \& Ma (2001) instead used power law models and
obtained a lower limit on the power law index, $\gamma > 0.8$ at
90\% confidence for a surface density $\Sigma \propto R^{-\gamma}$.
The question remained, though, whether these two classes of models
were realistic enough to provide robust, model independent
conclusions about lens galaxy centers. Mu\~noz et al.\ (2001)
introduced double power law models where the core region is allowed
to have a power law cusp whose index is independent of the density
profile at large radii. They found that the lack of a core image in
B1933+503 robustly implies $\gamma \gtrsim 0.6$ for that one
galaxy. Keeton (2001) studied core images statistically using
models with cuspy stellar components, treated as generalized
Hernquist (1990) models, embedded in dark matter halos. He found
that the models were inconsistent with the data, perhaps because
generalized Hernquist models may not accurately represent the
stellar components of galaxies on 10--100 pc scales.

The goal of this paper is to reconsider the core image problem
using more realistic models derived from nearby galaxies, and more
generally to discuss using core images as tools for studying the
centers of distant ($z \sim 0.3$--1) galaxies. Nearby early-type
galaxies have surface brightnesses that can be modeled as a Nuker
law (Lauer et al.\ 1996; Byun et al.\ 1996), and in \S2 we discuss
lensing by such galaxies. In \S3 we consider in a general way what
physical properties of lens galaxies determine core images
properties, or conversely what we can learn about lens galaxies by
studying core images. In \S4 we study in detail the core images
predicted by a sample of realistic lens galaxies. Finally, in \S5
we offer a summary and conclusions.  We assume the popular
$\Lambda$CDM cosmology with matter density $\Omega_M=0.3$,
cosmological constant $\Omega_\Lambda=0.7$, and Hubble constant
$H_0=75$ km s$^{-1}$ Mpc$^{-1}$.

\section{Nuker Law Lenses}

The lensing properties of a galaxy with projected mass density
$\Sigma$ are given by the lensing potential $\phi$ that satisfies
the two-dimensional Poisson equation
$\nabla^2\phi = 2\Sigma/\Sigma_{\rm cr}$. Here
$\Sigma_{\rm cr} = (c^2 D_{\rm os})/(4\pi G D_{\rm ol} D_{\rm os})$
is the critical surface density for lensing, where $D_{\rm ol}$,
$D_{\rm os}$, and $D_{\rm ls}$ are angular diameter distances
between the observer (``o''), the lens (``l''), and the source
(``s''). We consider a fiducial lensing situation with a lens
galaxy at redshift $z_l=0.5$ and a source at redshift $z_s=2$,
which yields a critical density of
$\Sigma_{\rm cr} = 2230\,M_{\odot}\mbox{ pc}^{-2}$ for our adopted
$\Lambda$CDM cosmology. The lensing deflection is given by
$\alpha=\nabla\phi$, and the magnification depends on the second
derivatives of $\phi$.  See the book by Schneider et al.\ (1992) for
a full discussion of lens theory.

Because lensing selects galaxies by mass, the sample of observed
lens galaxies is dominated by early-type galaxies. In many nearby
early-type galaxies the surface brightness distribution is well
described by a Nuker law (Lauer et al.\ 1995; Byun et al.\ 1996),
\begin{equation}
  I(R) = 2^{(\beta-\gamma)/\alpha}\, I_b\, \left({R \over r_b}\right)^{-\gamma}
    \left[1+\left({R \over r_b}\right)^{\alpha}\right]^{(\gamma-\beta)/\alpha} ,
\end{equation}
where $\gamma$ and $\beta$ are the inner and outer power law
indices, respectively, $r_b$ is the radius where the break in
the power law occurs, $\alpha$ gives the sharpness of the break,
and $I_b$ is the surface brightness at the break radius. If the
luminosity distribution has circular symmetry and the mass-to-light
ratio is $\Upsilon$, the lensing deflection is
\begin{eqnarray}
  \alpha_{\rm gal}(R) &=& {2^{1+(\beta-\gamma)/\alpha} \over 2-\gamma}\
    \kappa_b\, r_b\, \left({R \over r_b}\right)^{1-\gamma} \\
  && \times\
    {}_2 F_1\left[ {2-\gamma \over \alpha} , {\beta-\gamma \over \alpha} ,
    1 + {2-\gamma \over \alpha} , -\left({R \over r_b}\right)^{\alpha} \right] ,
    \nonumber
\end{eqnarray}
where $\kappa_b = \Upsilon I_b/\Sigma_{\rm cr}$ is the surface mass
density at the break radius in units of the critical density for
lensing, and ${}_2 F_1$ is the hypergeometric function. If the
stellar distribution has ellipsoidal symmetry, the lensing
deflection must be computed numerically using the formalism given
by, e.g., Schramm (1990).

Most galaxies contain central, supermassive black holes (e.g.,
Magorrian et al.\ 1998), so we consider adding them to the model. The
deflection from a black hole is $\alpha_{\rm bh}(R) = R_E^2/R$ where the
black hole's Einstein radius is (in angular units)
\begin{equation}
  R_E = \left[ {4 G M_{\bullet} \over c^2}\
    {D_{\rm ls} \over D_{\rm ol} D_{\rm os}} \right]^{1/2} .
\end{equation}
We normalize the black holes using the observed correlation between
the black hole mass $M_{\bullet}$ and the velocity dispersion $\sigma$
of the parent galaxy (Gebhardt et al.\ 2000; Merritt \& Ferrarese
2001). The net deflection is simply the sum of the deflections from
the Nuker component and the black hole.

\begin{figure}[t]
\centerline{\epsfxsize=8.5cm \epsfbox{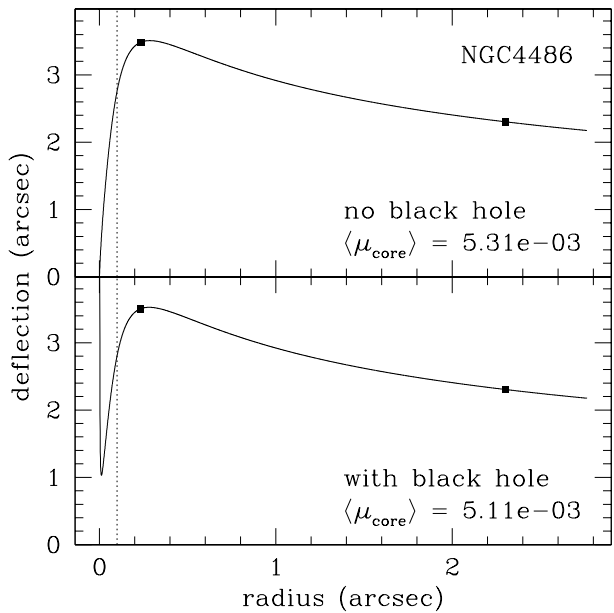}}
\caption{
The deflection profile for a mock lens galaxy obtained by taking
the nearby galaxy NGC4486 and moving it to redshift $z_l=0.5$. The
vertical dotted line indicates the Nuker break radius $r_b$. The
dots mark the critical radii $R_{\rm ein}$ and $R_{\rm rad}$. The
top panel shows the result for the Nuker galaxy alone, while the
bottom panel adds a central black hole normalized by the
$M_{\bullet}$--$\sigma$ relation from Gebhardt et al.\ (2000). The
mean core image magnification $\ma$ is computed with the method
presented in \S3.
}\label{fig:def}
\end{figure}

As an example we consider the nearby galaxy NGC4486, which has
Nuker parameters $\alpha=2.82$, $\beta=1.39$, and $\gamma=0.25$ and
is a fairly typical (albeit massive) example of the early-type
galaxies in the sample studied by Faber et al.\ (1997). We imagine
a lens galaxy obtained by moving NGC4486 to a typical lens redshift
$z_l=0.5$. The deflection profile $\alpha(R)$ for the resulting
lens is shown in \reffig{def}. The asymptotic behavior is
$\alpha(R) \propto R^{1-\gamma}$ for $R \ll r_b$ (if there is no
black hole), and $\alpha(R) \propto R^{1-\beta}$ for $R \gg r_b$.
Typical galaxies have $\gamma < 1$ and $\beta > 1$, so the
deflection is zero at the origin, rises to some finite peak, then
slowly declines. There are two important radii corresponding to the
``critical curves,'' or curves along which the lensing
magnification is infinite. The tangential critical curve lies at
the Einstein radius $R_{\rm ein}$, which is the solution to
$\alpha(R_{\rm ein}) = R_{\rm ein}$. The radial critical curve lies
at the radius $R_{\rm rad}$ which is the solution to $d\alpha/dR = 1$.
The radial critical curve maps to a caustic at
\begin{equation}
  u_{\rm max} = \alpha(R_{\rm rad}) - R_{\rm rad}\ . \label{eq:umax}
\end{equation}
This caustic bounds the multiply-imaged region; sources with
$u < u_{\rm max}$ are multiply-imaged, while sources with
$u > u_{\rm max}$ are not. The core images are always contained in the
region bounded by the radial critical curve. The radii $R_{\rm ein}$
and $R_{\rm rad}$ are marked in \reffig{def}.

If the galaxy has a steep central cusp with $\gamma>1$, the
deflection diverges at the origin and is a monotonically decreasing
function of radius. In this case, the radial critical curve does
not exist and the lens never produces a core image. If the galaxy
contains a central black hole, the black hole causes the deflection
to diverge at the origin and suppresses some of the core images
(Mao, Witt \& Koopmans 2001). However, the black hole changes the
deflection only at very small radii, so it has little effect on the
critical radii $R_{\rm ein}$ and $R_{\rm rad}$ or on the mean core
image magnification $\ma$ (see \reffig{def}). Adding a dark matter
halo to the lens galaxy would raise the outer deflection profile and
make it approximately flat, but it would have little effect on the
central deflection profile that determines the properties of the
core images unless it were much more centrally concentrated than
the light.

\begin{figure}[t]
\centerline{\epsfxsize=8.5cm \epsfbox{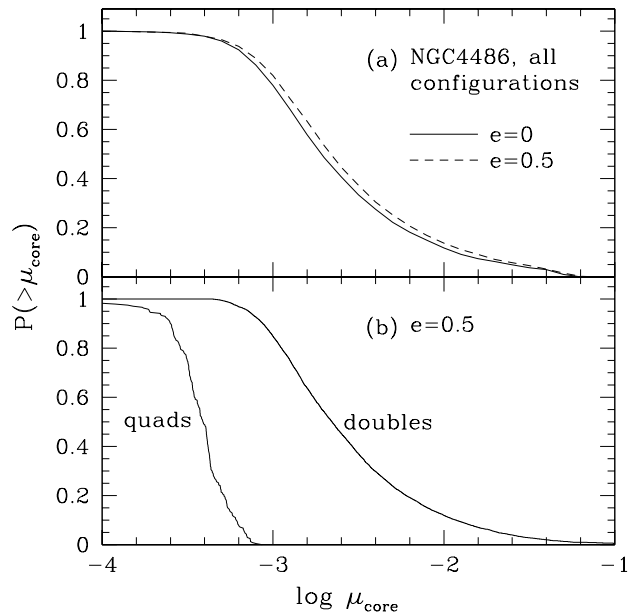}}
\caption{
Cumulative distributions of magnification factors for the core
images predicted by the NGC4486 mock lens galaxy.
(a) Total distributions plotted for a circular (solid line) or
flattened (dashed line) lens galaxy.
(b) Distributions plotted separately for lenses with two or four
bright images (doubles or quads), for a lens galaxy with
ellipticity $e=0.5$. There is more statistical uncertainty in the
curve for quads than for doubles, because of the 7917 random source
positions we examined only 238 of them corresponded to quads.
}\label{fig:ngc4486}
\end{figure}

To characterize the core images expected in this lens, we study the
core image magnification distribution. The distribution can be
computed fairly rapidly using inverse ray shooting (e.g., Kayser,
Refsdal \& Stabell 1986; Wambsganss 1997).  \reffig{ngc4486}a shows
the distribution for circular and flattened lens galaxies. The
distribution is broad, spanning more than two orders of magnitude,
with a median value $\mm = 0.0019$ and a mean value $\ma = 0.0053$
for the circular case. Making the galaxy flattened shifts the
magnification distribution to higher values; but even for ellipticity
$e=0.5$ the shift is only 0.05 dex, so the core image magnification
distribution is largely insensitive to ellipticity in the lens galaxy.
(We would see more of an effect if we described core images by their
flux ratio relative to the bright images, $\mm/\mu_{\rm bright}$,
because $\mu_{\rm bright}$ is quite sensitive to ellipticity.)

Ellipticity is important in one respect. If the lens galaxy is
non-spherical, some source positions correspond to lenses with two
bright images (doubles), while others correspond to lenses with
four bright images (quads). (See, e.g., Schneider et al.\ 1992.)
\reffig{ngc4486}b shows the core image magnification distributions
for quads and doubles separately. Because the source lie closer to
the origin for quads than for doubles, and the core image
magnification increases with the distance of the source from the
origin, quads tend to have smaller core image magnifications than
doubles. The quantitative details depend on the ellipticity, on any
external tidal perturbation (shear) that may affect the lens, and
on the stellar profile of the galaxy, but the general result is
robust: quads to tend have smaller $\mm$ than doubles.

The ellipticity effect combines with an observational selection
effect. Quads generally have larger total magnifications than
doubles, so the sources in quads tend to be intrinsically fainter
than the sources in doubles. Together, their fainter sources and
smaller core image magnifications suggest that quads will tend to
have fainter core images than doubles. The lack of core images in
observed quads should therefore be less surprising than in doubles.
Conversely, doubles should be better systems than quads for searching
for core images.

Thus, while ellipticity in the lens galaxy is important in
understanding differences between quads and doubles, it is not very
important in the overall distribution of core image magnifications.
In the remainder of the paper we therefore neglect ellipticity and
use circular lens galaxies.

\section{What Do Core Images Probe?}

In order to derive meaningful conclusions from observational
constraints on core images, it is important to understand how core
images depend on the physical properties of lens galaxies. (Merely
understanding the parameter dependencies in parametric lens models
does not allow strong physical conclusions.) Although the full
distributions of core images predicted for a particular lens galaxy
must be computed numerically, the mean magnification $\ma$ can be
studied analytically. Moreover, in \S4 we argue that this quantity
is actually a good way to characterize the distribution. Hence, in
this section we study $\ma$ in general terms.

Formally, we have
\begin{eqnarray}
  \ma &=& { \int_{\rm mult} \mm(\uu)\,d\uu \over \int_{\rm mult} d\uu }\ , \\
  &=& { \int_{\rm core} d\xx \over \int_{\rm mult} d\uu }\ . \label{eq:mu1}
\end{eqnarray}
The first line is simply the definition of the average, where the
integral extends over the multiply-imaged region of the source
plane. In the second line, the integral in the numerator extends
over the ``core'' region in the image plane, defined to be the
region within the radial critical curve; this equality holds
because $\mm = |\partial\xx / \partial\uu|$ is the Jacobian of the
transformation between the source and image planes. In words,
\refeq{mu1} says that the mean core image magnification is equal to
the area within the radial critical curve divided by the lensing
cross section (the area of the multiply-imaged region in the source
plane). For circularly symmetric lenses,
\begin{equation}
  \ma = \left( R_{\rm rad} / u_{\rm max} \right)^2 . \label{eq:mu2}
\end{equation}
For non-circular lenses, \reffig{ngc4486} suggests that this is
still a good approximation, because the $\mm$ distribution is not
terribly sensitive to ellipticity. These relations were first given
by Keeton (2001).

Now we focus on circularly symmetric lenses. Given the definition
of $u_{\rm max}$ from \refeq{umax}, we can write \refeq{mu2} as
\begin{eqnarray}
  \ma &=& \left[ \alpha(R_{\rm rad})/R_{\rm rad} - 1 \right]^{-2} ,
    \label{eq:mu3} \\
  &=& \left( \avg{\kappa}_{R_{\rm rad}} - 1 \right)^{-2} , \label{eq:mu4}
\end{eqnarray}
where $\avg{\kappa}_R$ is the mean surface density within radius
$R$, in units of the critical density for lensing. The second
equality holds because by the definition of the deflection,
\begin{equation}
  {\alpha(R) \over R}
  = {2 \over R^2} \int_{0}^{R} \xi\,\kappa(\xi)\,d\xi
  = \avg{\kappa}_R\ . \label{eq:kavg}
\end{equation}
Alternatively, returning to \refeq{mu3} and using the identities
\begin{eqnarray}
  {\alpha(R) \over R} + {d\alpha \over dR} &=& 2\,\kappa(R)\,,
    \label{eq:k} \\
  {d\alpha \over dR}\biggr|_{R_{\rm rad}} &=& 1\,,
\end{eqnarray}
we obtain
\begin{equation}
  \ma = {1 \over 4} \left[ \kappa(R_{\rm rad}) - 1 \right]^{-2} .
    \label{eq:mu5}
\end{equation}
\refEqs{mu4}{mu5} indicate that the mean core image magnification
is given very simply from either the surface mass density at the
radial critical curve $R_{\rm rad}$ or the mean surface mass density
within $R_{\rm rad}$.

What remains is to understand what physical properties of the
galaxy determine the radial critical curve $R_{\rm rad}$. Equating
\refeqs{mu4}{mu5}, we find that $R_{\rm rad}$ is the solution to
\begin{equation}
  2\,\kappa(R_{\rm rad}) = \avg{\kappa}_{R_{\rm rad}} + 1\,. \label{eq:k2}
\end{equation}
This relation implies that the radial critical curve is related to
the concentration of galaxy's (projected) mass distribution. We
expect $\kappa(R)$ to be a decreasing function, so
$\avg{\kappa}_{R} \ge \kappa(R)$ for all $R$. If $\kappa(R)$ is
steep (the mass is concentrated), then $\avg{\kappa}$ is large and
we must go to large $\kappa$ (small $R$) to satisfy \refeq{k2};
\refeq{mu5} then implies faint core images.\footnote{This result
also explains how adding a central black hole affects the mean core
image magnification. The black hole suppresses some core images,
reducing $\ma$ (Mao et al.\ 2001). In the language of our analysis,
the black hole increases $\avg{\kappa}_{R}$ without changing
$\kappa(R)$, so we must move to larger $\kappa$ (smaller $R$) to
keep \refeq{k2} satisfied.} Conversely, if $\kappa(R)$ is shallow
then \refeq{k2} is satisfied at smaller $\kappa$ (larger $R$), and
the core images are brighter.

These results can be demonstrated with two simple examples. First,
consider a softened isothermal sphere with surface mass density
$\kappa(R) = (b/2)(s^2+R^2)^{-1/2}$, where $s$ is a core radius and
$b$ is a scale radius that equals the Einstein radius in the case
$s=0$. The critical radii and mean core image magnification are
\begin{eqnarray}
  R_{\rm ein} &=& \left[ b(b-2s) \right]^{1/2} , \label{eq:iso1} \\
  R_{\rm rad} &=& {1 \over 2}\,\zeta (\xi+\zeta)^{1/2} (\xi-3\zeta)^{1/2} , \\
  u_{\rm max} &=& {1 \over 4}\,      (\xi+\zeta)^{1/2} (\xi-3\zeta)^{3/2} , \\
  \ma &=& \left({ 2 \zeta \over \xi-3\zeta }\right)^2 ,
\end{eqnarray}
where $\xi=(4b+s)^{1/2}$ and $\zeta=s^{1/2}$. Decreasing the core
radius reduces $\ma$, with $\ma \approx s/b$ for $s \ll b$.
Second, consider a power law density $\rho \propto r^{-\gamma}$ or
$\kappa \propto R^{1-\gamma}$, with $\gamma>1$ to ensure that
$\kappa(R)$ is a decreasing function. The critical radii and mean
core image magnification are
\begin{eqnarray}
  R_{\rm rad} &=& R_{\rm ein}\,(2-\gamma)^{1/(\gamma-1)} , \\
  u_{\rm max} &=& R_{\rm ein}\,(\gamma-1) (2-\gamma)^{(2-\gamma)/(\gamma-1)} , \\
  \ma &=& \left({ 2-\gamma \over \gamma-1 }\right)^2 . \label{eq:pow3}
\end{eqnarray}
Increasing $\gamma$ (making the profile steeper) decreases $\ma$.
These expressions are valid only for $1 < \gamma < 2$, because for
$\gamma>2$ the density profile is so steep that the radial critical
curve does not exist and the model does not produce core images.
Previous studies used models like these to understand the inverse
relation between the brightness of core images and the concentration
of the lens galaxy, but \refeqs{mu5}{k2} now give it in a general,
model independent form.

\section{Core Images in Realistic Galaxies}

\subsection{The sample}

To understand core images in realistic galaxies, we study models
constructed from a sample of observed galaxies. This approach
ensures that we examine the region of parameter space occupied by
real systems. We seek galaxies with well resolved luminosity
profiles, plus measured velocity dispersions as mass indicators.
Faber et al.\ (1997), Carollo et al.\ (1997), Carollo \& Stiavelli
(1998), and Ravindranath (2001) have published samples of nearby
early-type galaxies observed with high-resolution Hubble Space
Telescope imaging at optical or near-infrared
wavelengths.\footnote{Rest et al.\ (2001) give a similar sample,
but without velocity dispersions.} Together the samples comprise 73
distinct galaxies with both Nuker law fits and velocity
dispersions, which are summarized in \reftab{sample}. The different
samples use different passbands: V for the Faber sample, H for the
Ravindranath sample, and V and I for the Carollo sample. However,
we can check for wavelength dependence and other systematic effects
because some of the galaxies appear in more than one sample: six
galaxies in both the Faber and Ravindranath samples, five galaxies
in both the Faber and Carollo samples, six galaxies in both the
Ravindranath and Carollo samples, and ten galaxies with multiple
passbands in the Carollo sample. (No galaxies appear in all three
samples.)

We use this sample to construct mock lens galaxies by moving each
galaxy to a typical lens redshift $z_l=0.5$, and assuming a typical
source redshift $z_s=2$. Following Faber et al.\ (1997), we compute
the mass-to-light ratio for each galaxy using a spherical and
isotropic dynamical model. Faber et al.\ (1997) give mass-to-light
ratios for their sample, providing a check for our calculations. We
then compute the lensing properties of the mock lens galaxies,
which are summarized in \reftab{sample}.

As discussed in \S2 we focus on circularly symmetric galaxies
because ellipticity has little effect on the overall distribution
of core image magnifications. We consider only the Nuker components
of the galaxies, neglecting any nuclear components that may be
indicated by fits to the surface brightness distribution. This is
a conservative approach to our problem, because any additional
mass concentration would decrease the predicted core image
magnifications and bring the models closer to agreement with the
data. We also neglect dark matter halos. Dark matter does not
appear to be dynamically important in the inner 5--10 kpc of
elliptical galaxies (e.g., Gerhard et al.\ 2001). While it is
important for lensing (because lensing depends on the projected
mass; e.g., Treu \& Koopmans 2002), dark matter would have little
effect on the projected mass distributions on the $\lesssim$200 pc
scales important for core images unless it were substantially more
concentrated than the light.

\begin{figure}[t]
\centerline{\epsfxsize=8.5cm \epsfbox{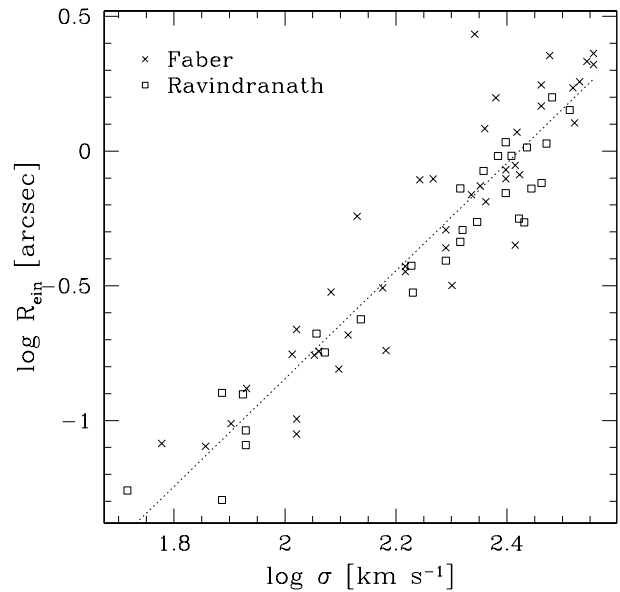}}
\caption{
Relation between the velocity dispersion $\sigma$ and Einstein
radius $R_{\rm ein}$ for the mock lens galaxies. Crosses and boxes
indicate galaxies in the Faber and Ravindranath samples,
respectively. The dotted line shows the relation
$R_{\rm ein} \propto \sigma^2$; the fitted zero point is $-4.84$
(in log units), compared with a predicted value of $-4.74$ for
comparable Singular Isothermal Spheres.
}\label{fig:sig_re}
\end{figure}

We can check that our mock lenses are reasonable in several ways.
First, we compute the Einstein radius for each galaxy and compare
it to the velocity dispersion in \reffig{sig_re}; this tests
whether the lensing and dynamical masses are consistent. For
comparison, a simple Singular Isothermal Sphere (SIS) lens has the
scaling $R_{\rm ein} \propto \sigma^2$, with a zero point of $-4.74$
(in log units) for $z_l=0.5$ and $z_s=2$ in our adopted $\Lambda$CDM
cosmology (see Schneider et al.\ 1992). The mock galaxies are
consistent with this scaling and a fitted zero point of $-4.84$,
although with a scatter of $0.16$ dex. The fact that the mock
galaxies lie on average $0.1$ dex below the expected SIS relation
may be related to our neglect of dark matter halos. Still, the
dynamical and lensing properties are related in a sensible way
suggesting that the mock lens galaxies are not unreasonable.

\begin{figure}[t]
\centerline{\epsfxsize=8.5cm \epsfbox{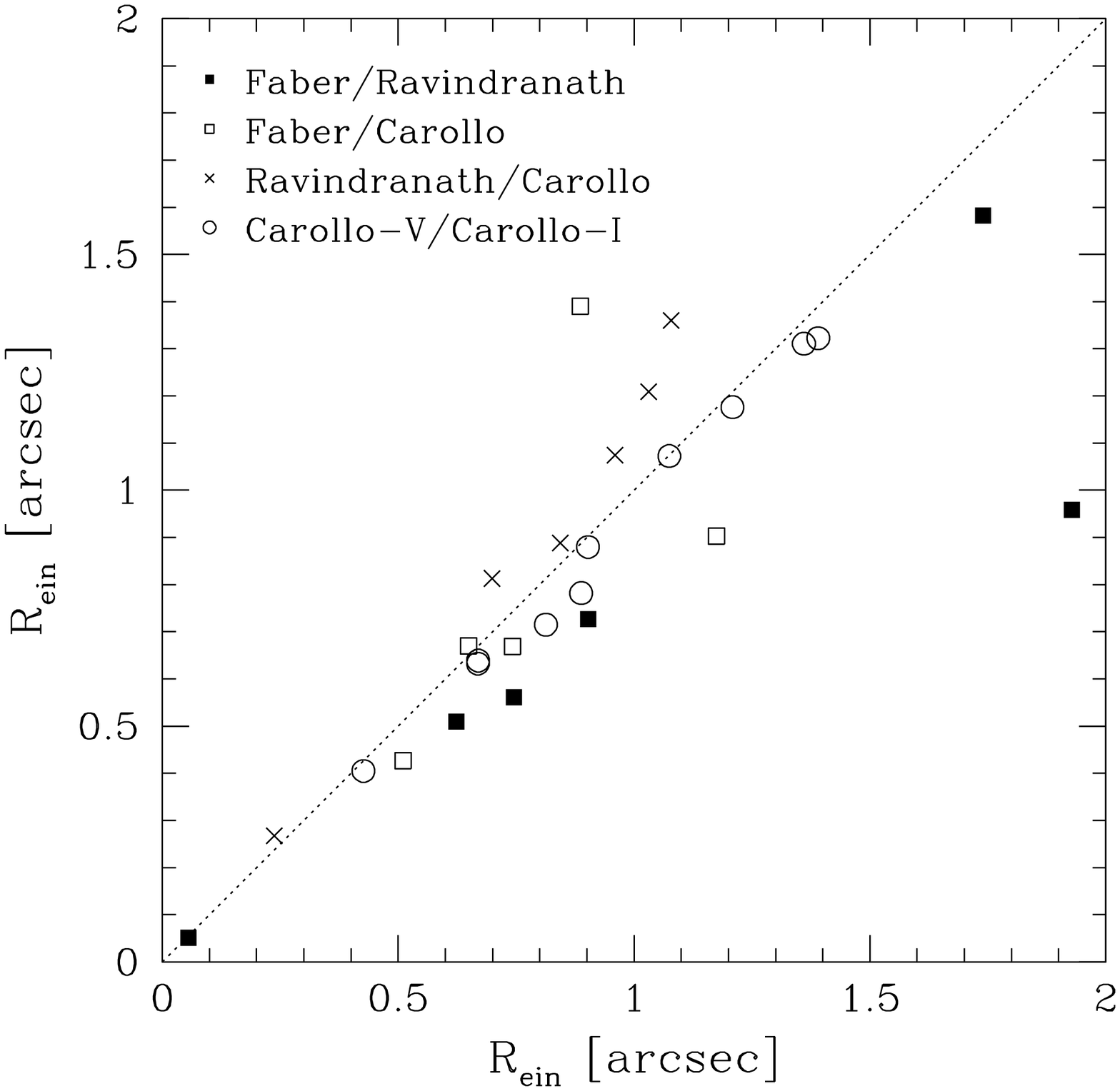}}
\centerline{\epsfxsize=8.5cm \epsfbox{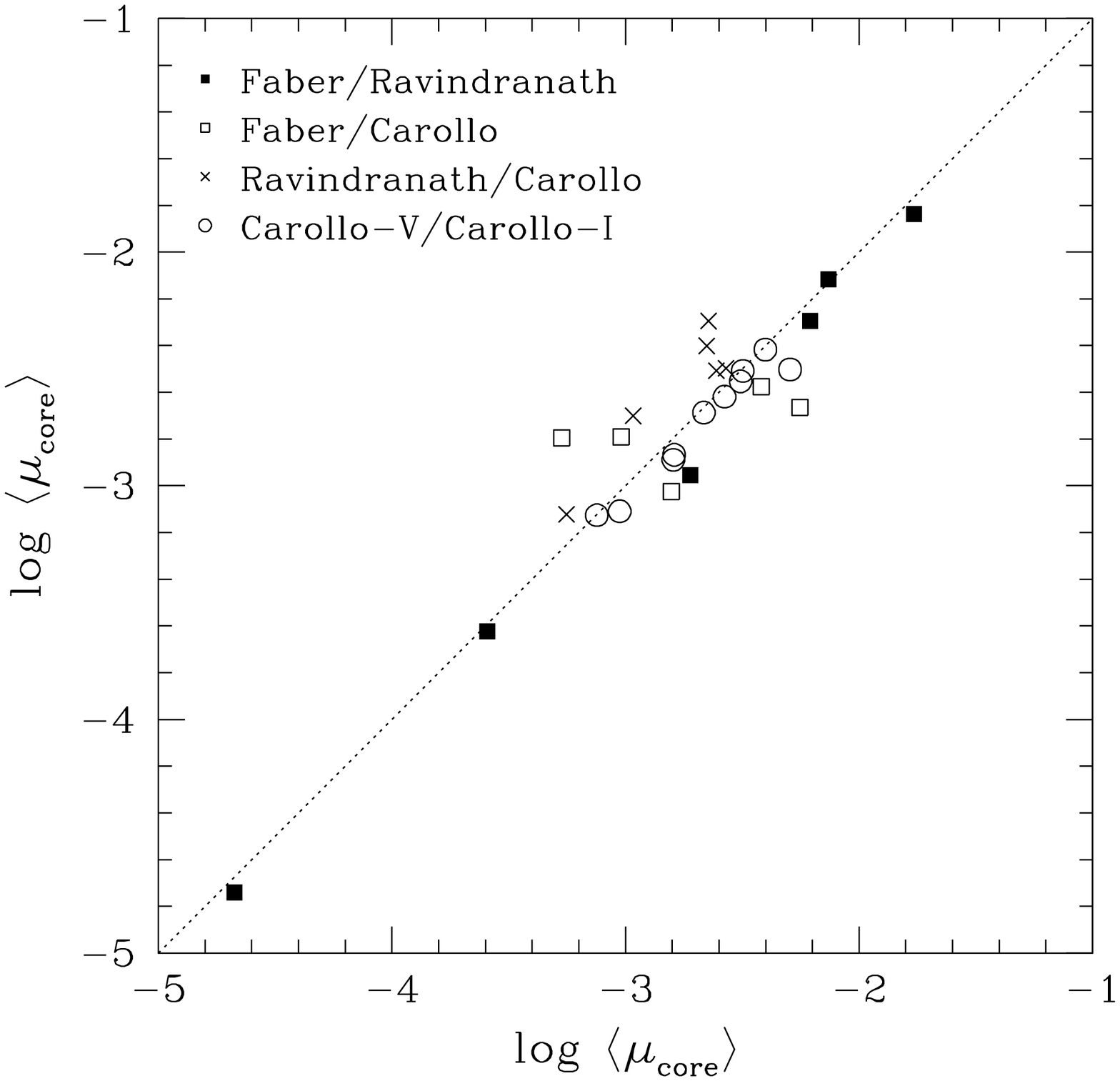}}
\caption{
A comparison of the Einstein radii (top) and mean core image
magnifications (bottom) for galaxies that appear in more than one
of the original samples. Filled boxes indicate galaxies in both the
Faber and Ravindranath samples, open boxes the Faber and Carollo
samples, and crosses the Ravindranath and Carollo samples. Open
circles indicate galaxies in both the V-band and I-band Carollo
samples.
}\label{fig:overlap}
\end{figure}

As a second check, we consider the galaxies that appear in more
than one of the original samples. For example, for each galaxy that
was observed and modeled by both Faber et al.\ (1997) and
Ravindranath et al.\ (2001), we compute the lensing properties
using both models and compare them. In this way we test whether the
use of different modeling techniques and different passbands
affects our lensing results. \reffig{overlap} compares the Einstein
radii and mean core image magnifications for all of the duplicate
galaxies. There is fair agreement in the Einstein radii, although
with some scatter because different studies find somewhat different
values of the outer slope $\beta$; the main outlier is NGC524,
where dust is known to affect the luminosity profile at optical
wavelengths (Lauer et al.\ 1995; Ravindranath et al.\ 2001). There
is good agreement in the mean core image magnification, so our
conclusions about the properties of core images are not sensitive
to whose data we use. For the remainder of the paper, we adopt as
our main sample all of the Ravindranath galaxies plus the Faber
galaxies that are not in the Ravindranath sample.

\subsection{A plethora of core images}

\begin{figure}[t]
\centerline{\epsfxsize=8.5cm \epsfbox{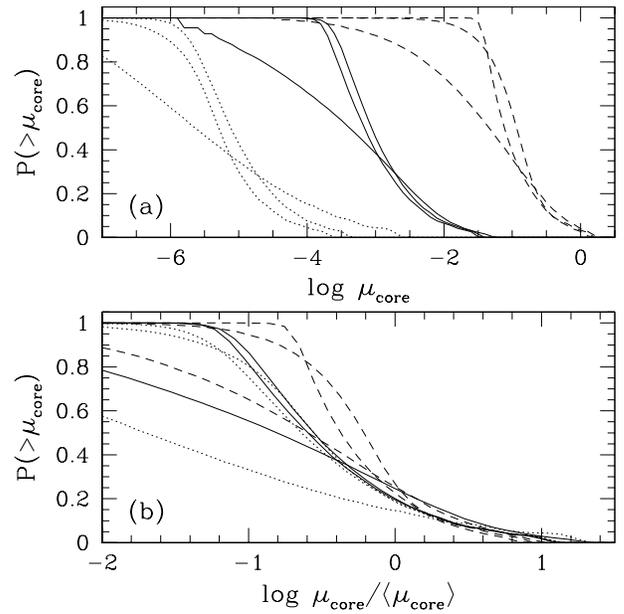}}
\caption{
Cumulative core image magnification distributions for nine of the
mock lens galaxies: the three with the lowest non-zero values for
$\ma$ (dotted lines; NGC221, NGC3377, NGC4621); the three with
values for $\ma$ at the median of the sample (solid lines; NGC4570,
NGC5982, NGC4649); and the three with the largest values for $\ma$
(dashed lines; NGC4239, NGC6166, NGC5273).
(a) Distributions of the core image magnification $\mm$.
(b) Distributions of the normalized magnification $\mm/\ma$.
}\label{fig:mdist}
\end{figure}

\reffig{mdist}a shows the core image magnification distributions
for nine of the mock lens galaxies to illustrate the range of
effects. Each mock lens galaxy has a distribution of core image
magnifications that spans some two orders of magnitude, and the
complete set of mock lenses spans six orders of magnitude in $\mm$.
Realistic lens galaxies predict a remarkably wide range of core
images.

Studying the full $\mm$ distribution for each mock lens galaxy is
impractical, so we would like to describe each galaxy by a single
characteristic quantity. We propose to use the mean core image
magnification $\ma$ as a good characteristic value, partly because
this quantity is easy to compute (see \S3), and partly because
\reffig{mdist}b suggests that it is indeed characteristic of the
overall distribution. Specifically, when plotted in terms of the
normalized quantity $\mm/\ma$ the distributions for a wide range of
galaxies all lie on top of each other. Although there are
differences in the shapes of the distributions at the faint end,
the distributions at the bright end are remarkably similar. Hence,
we believe that $\ma$ is a useful characterization of the
distribution of core image magnifications for a particular galaxy,
especially at the bright end.

\begin{figure}[t]
\centerline{\epsfxsize=8.5cm \epsfbox{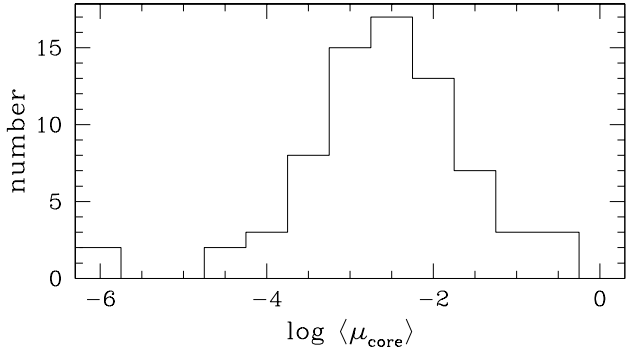}}
\caption{
Histogram of the mean core image magnifications for the 73 mock
lens galaxies. Galaxies that do not produce core images are
arbitrarily placed at $\log\ma=-6$.
}\label{fig:mhist}
\end{figure}

\reffig{mhist} shows a histogram of the $\ma$ values for the 73
mock lens galaxies. (The values are given in \reftab{sample}.) Note
that the histogram is intended only to show the broad range of core
image properties in our sample; it should not be interpreted as a
global distribution of $\ma$ values, because our sample is not a
statistical sample of galaxies. Still, the histogram is instructive
in illustrating that the mean values $\ma$ span more than four
orders of magnitude, from $\ma=1.8\E{-5}$ for NGC221 to $\ma=0.19$
for NGC5273 --- plus two galaxies that never produce core images
(NGC1172 and NGC4742). Again we see the wide range of core image
properties in realistic lens galaxies; some lenses should have
bright, detectable core images, while others should have core
images that are essentially invisible.

\begin{figure}[t]
\centerline{\epsfxsize=8.5cm \epsfbox{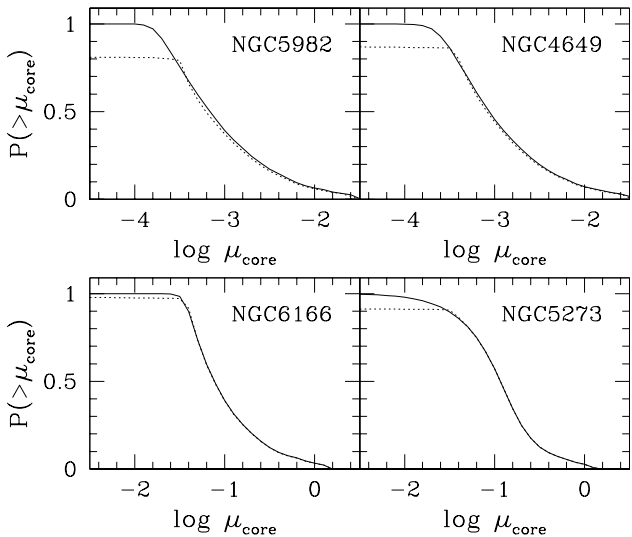}}
\caption{
Core image magnification distributions for galaxies with (dotted
lines) and without (solid lines) central supermassive black holes,
for four of the galaxies from \reffig{mdist}. Black holes can
suppress core images (see Mao et al.\ 2001), but only at the faint
end of the core image distribution.
}\label{fig:mdistBH}
\end{figure}

In principle the central supermassive black holes that are common
in galaxies can suppress core images (Mao et al.\ 2001), but in
practice they have little effect. \reftab{sample} gives the values
for $\ma$ when the galaxies contain black holes normalized by the
$M_{\bullet}$--$\sigma$ relations measured by Gebhardt et al.\
(2000) and Merritt \& Ferrarese (2001). In a few cases the black
hole modifies the central potential enough to erase all core images
(NGC221, NGC3115, NGC3377, NGC3900, NGC4464, NGC4467, NGC4621, and
NGC5838). But all of these cases have $\ma \le 0.001$ even without
a black hole, which suggests that adding a black hole can erase all
core images only if the core images are faint to begin with. In the
remaining cases the black hole reduces $\ma$ by only about 0.1 dex.
The black hole does suppress some core images, but only at the {\it
faint\/} end of the distribution, as shown in \reffig{mdistBH}. In
other words, black holes have little effect on the core images in
realistic lens galaxies, especially at the bright end of the core
image distribution. Black holes therefore fail to resolve the core
image paradox. These conclusions are based on local measurements of
the $M_{\bullet}$--$\sigma$ relation, which may not hold at higher
redshifts; but black holes would have to be substantially more
massive (relative to their parent galaxies) at $z \sim 0.5$ than at
$z=0$ in order to change our conclusions. Another possibility is
that binary black holes are common in galaxies at $z \sim 0.5$. In
principle binary black holes can be more effective than single
black holes at suppressing core images, but for realistic binary
parameters the differences are small and binary black holes do not
strongly suppress core images (Keeton \& Zhao, in prep.). In the
remainder of the paper we neglect black holes.

\begin{figure}[t]
\centerline{\epsfxsize=8.5cm \epsfbox{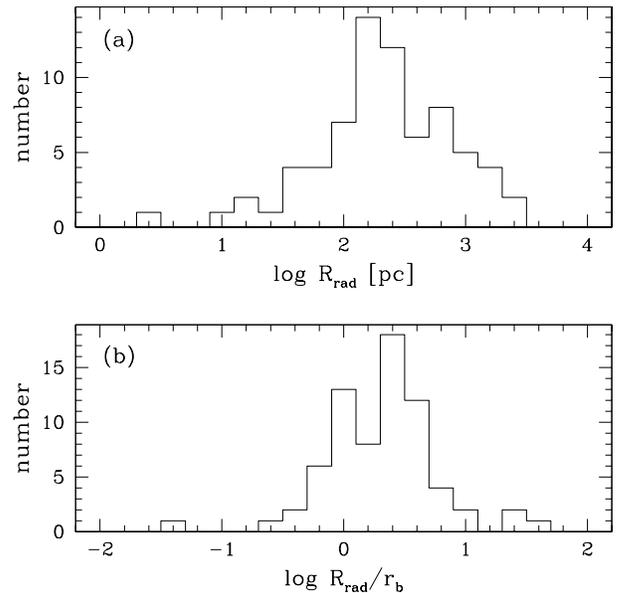}}
\caption{
Histograms of the lensing critical radius $R_{\rm rad}$, which bounds
the core image region.
(a) The critical radius in parsecs.
(b) The ratio of the critical radius to the Nuker break radius
$r_b$.
Galaxies that lack a critical radius $R_{\rm rad}$ are not included.
}\label{fig:rhist}
\end{figure}

We now seek to understand the diversity of core image properties
in the mock lens galaxies, and to identify the galaxy properties
that affect the core images. First, we examine the region of the
galaxy that is probed by core images. Recall that the core images
are always contained within the inner lensing critical curve, which
has radius $R_{\rm rad}$ (see \S2). \reffig{rhist} shows that the core
image region has a characteristic scale $\sim$200 pc (the median
value of $R_{\rm rad}$ for the sample), and is comparable to or
slightly larger than the Nuker break radius. Thus, core images
probe the inner tens to hundreds of parsecs at the centers of
galaxies.

\begin{figure*}[t]
\centerline{\epsfxsize=16.0cm \epsfbox{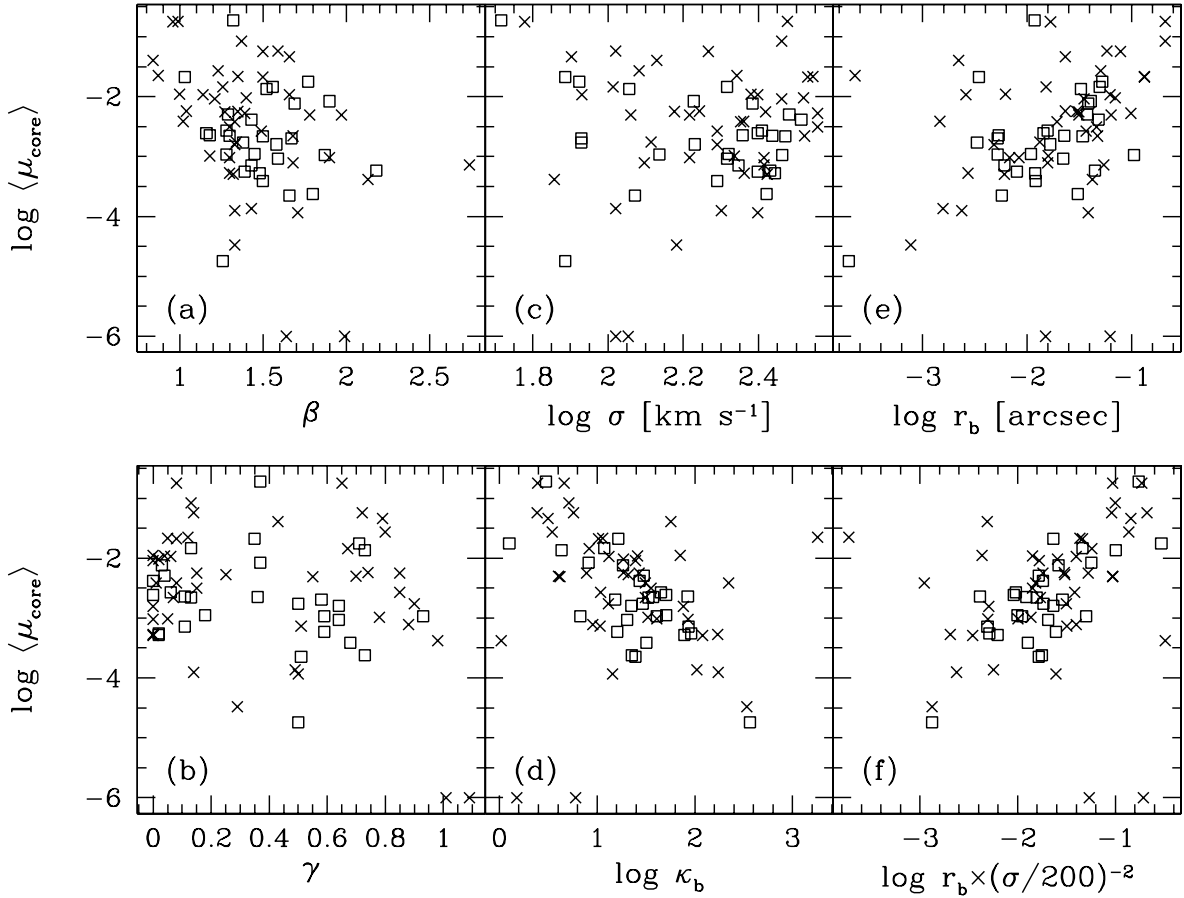}}
\caption{
Scatter plots of the mean core image magnifications $\ma$ versus
various galaxy properties. Crosses and boxes indicate galaxies in
the Faber and Ravindranath samples, respectively. Galaxies that do
not produce core images are arbitrarily placed at $\log\ma=-6$.
}\label{fig:scatt}
\end{figure*}

Next, we consider how the mean core image magnification $\ma$
depends on various properties of the galaxies, as shown in
\reffig{scatt}. The interesting general result is that no single
property of a galaxy strongly determines its core image properties.
There is no simple relation between $\ma$ and the outer and inner
Nuker power law indices $\beta$ and $\gamma$
(Figs.~\ref{fig:scatt}a and \ref{fig:scatt}b). This result makes
sense in combination with \reffig{rhist}: the lensing critical
radius $R_{\rm rad}$ is often comparable to the Nuker break radius
$r_b$, so we are in a regime where the galaxy cannot be described
as a simple power law, and $\ma$ is not dominated by either $\beta$
or $\gamma$ separately. The distinction between ``core'' and
``power law'' galaxies seen in their luminosity profiles and
dynamics (Faber et al.\ 1997) does not appear to carry over into
lensing and core images. There is likewise no simple relation between
$\ma$ and galaxy mass (as represented by velocity dispersion;
\reffig{scatt}c); galaxies with a given $\sigma$ have $\ma$ values
that range over some three orders of magnitude or more. Dwarf
galaxies are not systematically more or less likely than giant
galaxies to produce bright core images.

Based on the arguments in \S3 we expect a connection between the
core image properties and some measure of the concentration of the
mass distribution. There is a trend between $\ma$ and the surface
density at the break radius ($\kappa_b$, \reffig{scatt}d): galaxies
with higher break densities tend to predict fainter core images.
There is also a general trend between with the Nuker break radius
$r_b$ (\reffig{scatt}d): galaxies with smaller break radii, and
hence smaller core regions, tend to predict fainter core images.
But this trend is neither strong nor tight, because by itself $r_b$
cannot distinguish between galaxies that are large and highly
concentrated and those that are less concentrated but intrinsically
small. One way to remove this effect is to normalize the break
radius using the Einstein radius as a measure of the global scale
of the galaxy; because $R_{\rm ein} \propto \sigma^2$, we actually use
$r_b/\sigma^2$ (\reffig{scatt}f). These three trends all indicate
that there is indeed some connection between the core images and
the concentration of the galaxy such that more concentrated
galaxies tend to predict fainter core images. However, the trends
have significant scatter and thus are not highly predictive. In
other words, there does not appear to be a simple measure of a
galaxy's concentration that strongly determines its core image
properties.

To summarize, realistic lens galaxies have an extremely wide range
of core image properties; some should have bright, detectable core
images, while others should have core images that are very faint
or absent altogether. A galaxy's core image properties are related
to its mass concentration, but there does not appear to be any simple
measure of concentration that yields a clean prediction of the core
image properties. The complication is that in Nuker law lenses the
lensing critical radius $R_{\rm rad}$ tends to be comparable to the break
radius $r_b$, which means that all of the Nuker parameters affect
the core image properties. This conclusion has a somewhat surprising
corollary. We might expect that the distinction between ``core'' and
``power law'' galaxies, with their different luminosity profiles,
would be obvious in their core image properties, but it is not.

\subsection{Should we see core images?}

We can now re-evaluate the core image problem in terms of our
expectations for realistic lens galaxies. Although our sample is not
a proper statistical sample of galaxies, if we assume that it does
at least represent the range of realistic galaxy properties then we
can consider whether the lack of observed core images is surprising
or not. The main issue is to understand what types of lens galaxies
produce bright core images and whether we should expect to find
many lenses from such galaxies. There are three galaxies in the
sample with $\ma > 0.1$: two are dwarf galaxies (NGC4239 and
NGC5273), while the third is a giant elliptical at the center of a
cluster (NGC6166). Because lensing selects galaxies by mass it
tends to select against dwarf galaxies, so even if some dwarf
galaxies are good at producing core images they are unlikely to
produce lenses in the first place. As for NGC6166, it is unusual
for being a brightest cluster galaxy as well as the most distant
(120 Mpc) and least concentrated (lowest surface brightness and
largest break radius) galaxy in the Faber et al.\ (1997) sample.
Thus, it appears to be an atypical galaxy drawn from the tail of
the galaxy population. Statistically, lensing is not likely to
select rare galaxies to be lens galaxies.

\begin{figure}[t]
\centerline{\epsfxsize=8.5cm \epsfbox{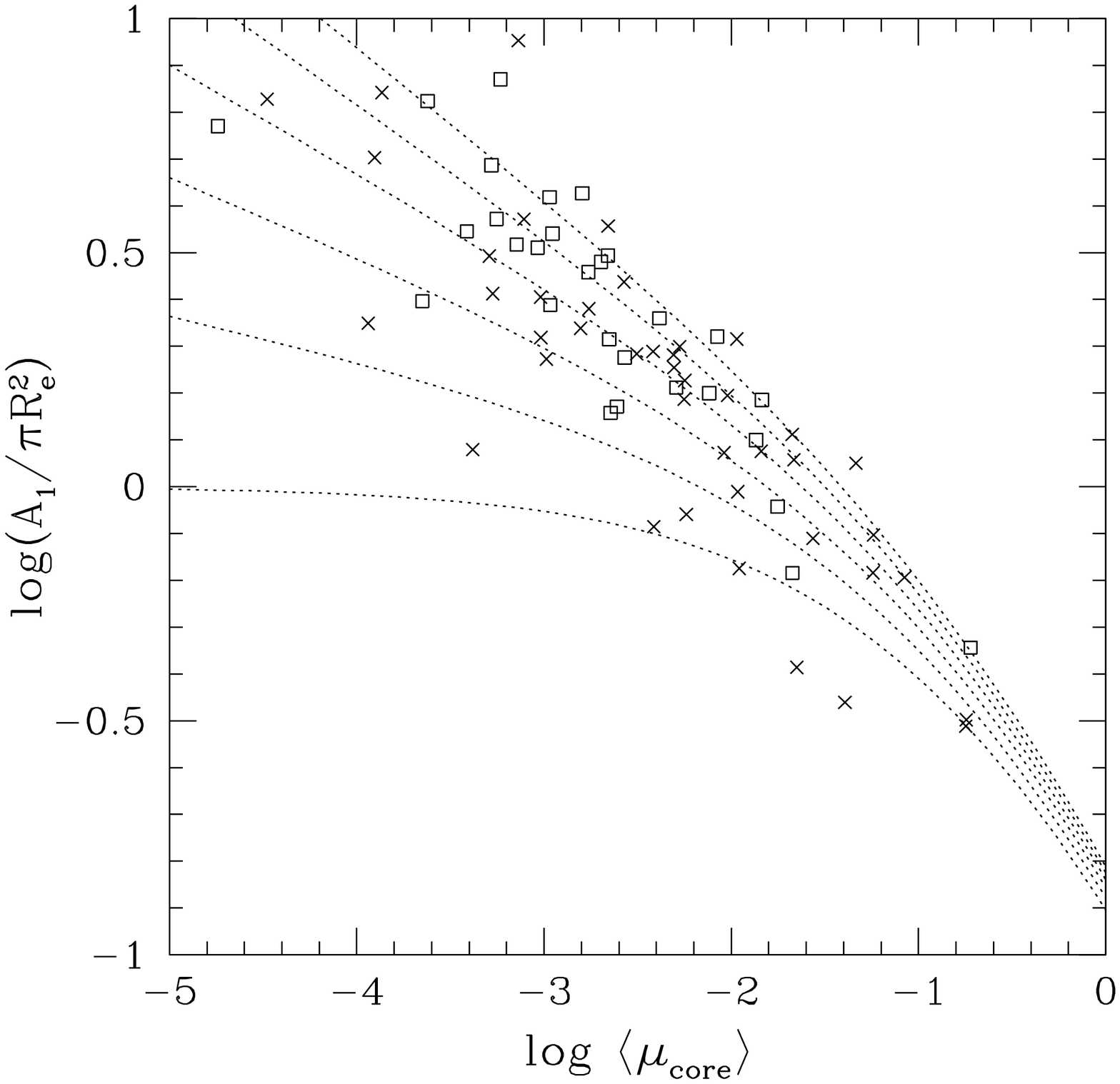}}
\centerline{\epsfxsize=8.5cm \epsfbox{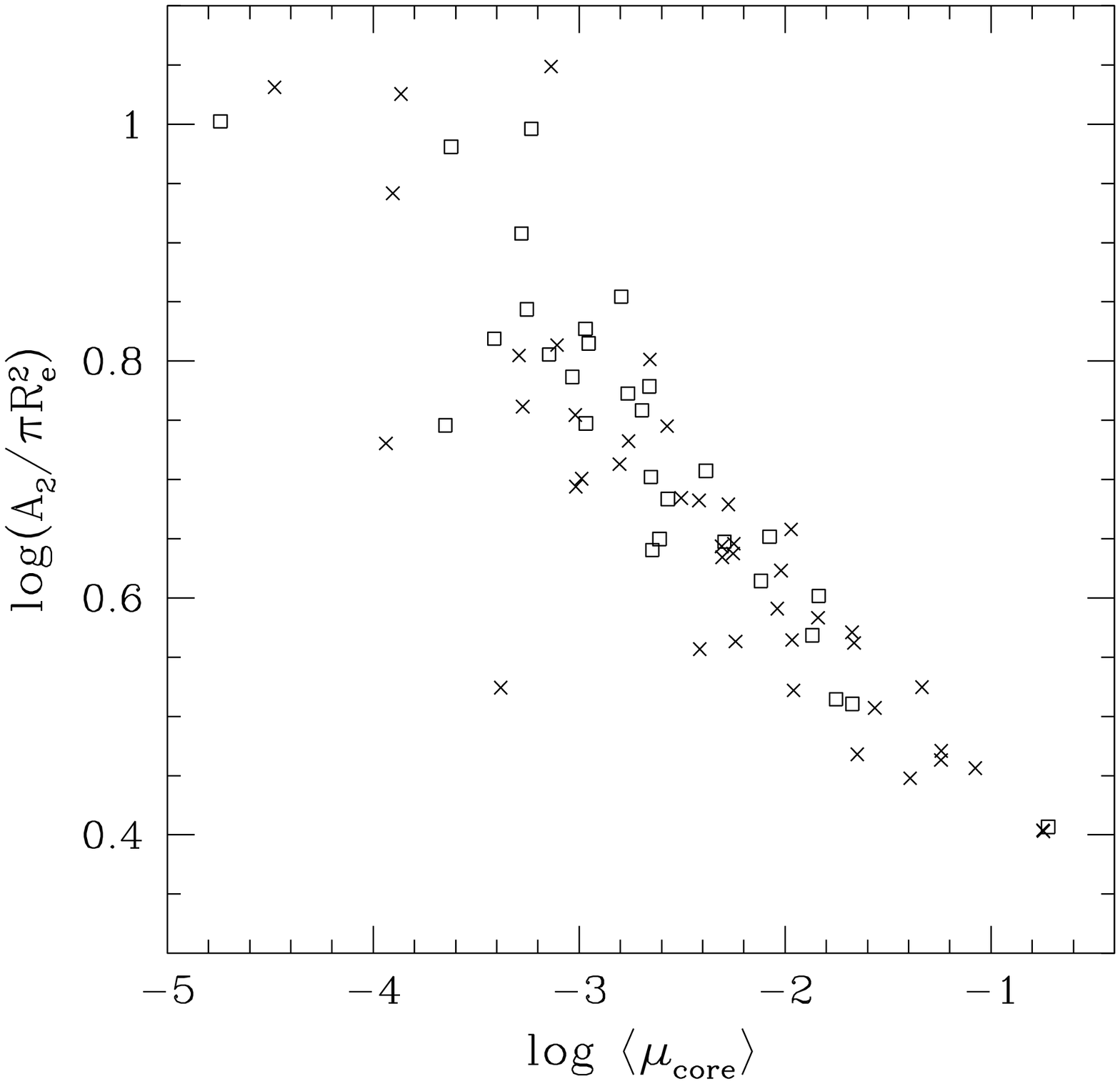}}
\caption{
Scatter plots of the lensing cross section versus the core image
magnification. The cross sections are normalized by the area within
the Einstein ring, $\pi R_{\rm ein}^2$.
(Top) $A_1$ is the simple lensing cross section, or the area of the
multiply-imaged region in the source plane. The curves show results
for simple softened power law lens models
$\Sigma \propto (s^2+R^2)^{-\beta}$ with $\beta = 1.0$, 1.1, 1.2,
1.3, 1.4, and 1.5 from bottom to top.
(Bottom) $A_2$ is the lensing cross section corrected for
magnification bias.
}\label{fig:norm}
\end{figure}

Lensing selection effects are important for core images in another
way. Consider a set of galaxies that are similar to each other,
specifically a set of galaxies with the same Einstein radius.
Within this set, are galaxies that predict bright core images any
more or less likely to be selected for lensing than galaxies that
predict faint core images? \reffig{norm} shows the lensing cross
section, normalized by the area within the Einstein ring, versus
$\ma$. There is a clear decrease in the normalized cross section
with increasing $\ma$, and it does not depend on whether
magnification bias is included or omitted. The trend for the mock
lens galaxies agrees well with analytic predictions for simple
softened power law lens models. At fixed Einstein radius, then,
galaxies that predict bright core images have smaller lensing cross
sections than galaxies that predict faint core images, so they are
less likely to be selected for lensing. There is an intrinsic bias
against lenses with bright core images.

Thus, the types of galaxies that can produce bright core images are
probably not common, and they have small lensing cross sections
relative to galaxies that are comparably massive but produce faint
core images. The two effects combine to suggest that bright core
images are not likely to be prevalent in observed lens samples.

\subsection{Comparison with data}

The observational constraints on core images take the form of upper
limits on the core image flux $f_{\rm core}$ in observed lenses. The
magnification cannot be directly constrained because the intrinsic
flux of the source is unknown. One way around this problem is to
use the flux ratio of the core image to the brightest image,
because the source flux factors out to leave
$f_{\rm core}/f_{\rm bright} = \mm/\mu_{\rm bright}$. This approach
requires explicitly solving the lens equation to find all images,
and thus is valuable for applications like lens modeling where the
lens equation must be solved anyway (e.g., Mu\~noz et al.\ 2001).
While it can be used for statistical analyses (e.g., Rusin \& Ma
2001; Keeton 2001), it eliminates technical simplifications like
the ability to compute $\ma$ rapidly without solving the lens
equation (but see Evans \& Hunter 2002 for a different kind of
technical simplification). It also makes the conclusions sensitive
to quantities like ellipticity in the lens galaxy, because $\mm$ is
fairly insensitive to ellipticity but $\mu_{\rm bright}$ is not.

\begin{figure}[t]
\centerline{\epsfxsize=8.5cm \epsfbox{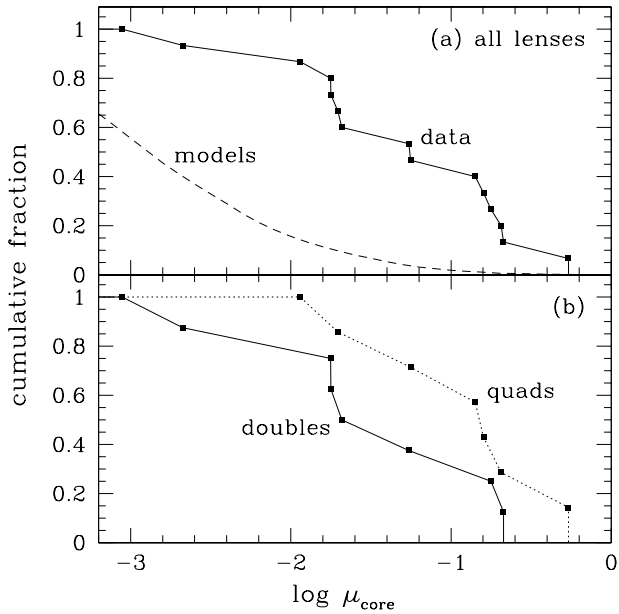}}
\caption{
Upper limits on core image magnifications for 15 observed radio
lenses (Norbury et al., in prep.). The points show 5$\sigma$ upper
limits on $\mm$.
(a) The data for all lenses. The dashed curve shows the net $\mm$
distribution predicted by a weighted sum of the model galaxies (see
text).
(b) The data for doubles and quads shown separately.
}\label{fig:data}
\end{figure}

An alternate approach is to fit a lens model to the observed images
to constrain the source flux, and then combine the inferred source
flux with the limits on $f_{\rm core}$ to put limits on $\mm$. This
approach should be robust because the model properties that
determine the source flux from the observed images (the enclosed
mass on 3--10 kpc scales) decouple from the properties that affect
core images (the density profile on $\lesssim$200 pc scales). With
this approach, Norbury et al.\ (2002) obtain upper limits on $\mm$
for 15 radio lenses from the Cosmic Lens All-Sky Survey, as shown
in \reffig{data}.

For an accurate comparison we need a prediction of the overall
$\mm$ distribution from the models (not merely the set of $\ma$
values for the sample galaxies, as in \reffig{mhist}). A proper
prediction is impossible because, again, our sample is not a
statistical sample of galaxies. Nevertheless, for the sake of
comparison we na{\"\i}vely combine our sample by summing all of the
$\mm$ distributions (a few of which are shown in \reffig{mdist}),
weighting each galaxy by its lensing cross section. The result is
shown as the dashed curve in \reffig{data}. It is interesting to
see that the model predictions lie well below the limits from the
data. The model does not predict that core images should be brighter
than observed. In fact, it suggests that the observational
sensitivity may need to improve by more than an order of magnitude
before detections of core images become common. Despite concerns
that the model sample is not statistically complete, we believe
that the general conclusion is reliable. Galaxies that produce
bright core images would have to be substantially more common in
the universe than in our sample in order to make the model
predictions inconsistent with the current observational data.

Evans \& Hunter (2002) reach similar conclusions from an analysis
of softened power law potential models. They argue that the break
radii in the galaxies observed by Faber et al.\ (1997) are small
enough to make core images faint even if the galaxies have simple
finite-density cores. Although we believe our models to be more
realistic because they allow more general cores and are constructed
directly from the fitted profiles, it is reassuring that the
conclusions are consistent.

We previously found that Hernquist model galaxies predict core
images to be more common than observed (Keeton 2001). Like a Nuker
law, a Hernquist model has a steep outer profile that smoothly
changes to a shallow central cusp, but with a larger transition
radius. Our new models should be more realistic because they are
based on fits to the luminosity profiles of observed galaxies,
which have small transition radii. The smaller break radii mean
higher central densities and thus fainter predicted core images.

Another interesting result from the data is that the upper limits
from quads are weaker than the upper limits from doubles
(\reffig{data}b). The flux limits for the various lenses are
comparable, but the quads tend to have fainter sources.
Combining this result with our prediction that quads will tend
to have fainter core images than doubles (\reffig{ngc4486}b),
we conclude that the lack of core images in observed quad lenses
is no surprise at all.

\section{Conclusions}

Core images in strong gravitational lens systems provide a unique
probe of the centers of galaxies at redshifts $z \sim 0.2$--1. The
brightnesses of core images are determined by the density profiles
of galaxies inside $\lesssim$200 pc. The lack of core images in
observed lenses, especially in radio lenses, sets strong lower
limits on the central densities of the lens galaxies (e.g.,
Wallington \& Narayan 1993; Mu\~noz et al.\ 2001).

The mapping between core images and galaxy centers can be studied
in two directions. In the forward problem knowledge of galaxy
centers is used to make predictions about core images. Based on the
first lens models drawn directly from the resolved stellar mass
distributions of nearby early-type galaxies, we predict that real
galaxies should produce a remarkably wide range of core images. Some
should have bright core images (magnification $\mm \gtrsim 0.1$),
while many others will have core images that are faint
($\mm \lesssim 0.001$) or absent altogether.

Qualitatively, more concentrated galaxies produce fainter core
images. Quantitatively, however, there does not seem to be a simple
predictive relation between observed galaxy properties and core
images. Lensing is biased against galaxies with bright core images,
because they have smaller cross sections than comparable galaxies
with faint core images. Four-image lenses should tend to have fainter
core images than two-image lenses, because in quads the source is
always close to the center of the lens galaxy where the core image
magnification is low. Supermassive black holes in the centers of
galaxies can suppress faint core images, but they have little effect
on bright core images or on the mean magnification.

The connection between core images and galaxy centers can also be
studied in the inverse problem, where the analysis of core image
data yields constraints on the centers of lens galaxies. Previous
work placed limits on the core radius or on the logarithmic slope
of the density (e.g., Wallington \& Narayan 1993; Rusin \& Ma 2001),
but it was not clear how model dependent those constraints were. We
obtain a general statement of the connection between the density
profile and core images: the mean core image magnification $\ma$ is
inversely related to the density at the lensing critical radius
$R_{\rm rad}$ (eq.~\ref{eq:mu5}); and this critical radius is
determined by the shape of the density profile (eq.~\ref{eq:k2}),
with more concentrated galaxies corresponding to smaller critical
radii and fainter core images. Unfortunately, neither our general
formalism nor our Nuker law lenses suggest any simple, model
independent measure of the mass concentration that determines the
core image properties. The interpretation of core image data will
therefore continue to rely on detailed models of individual lenses.
The model dependence can be held in check, though, by using general
models like the Nuker law or the cuspy lenses introduced by Mu\~noz
et al.\ (2001), as opposed to overly simple flat core or pure power
law models.

We conclude that in many cases the stellar mass in lens galaxies is
probably concentrated enough to render core images faint (also see
Evans \& Hunter 2002). This is not to say that bright core images
cannot exist --- certainly there are realistic galaxies that predict
bright core images, and the probability that they are selected for
lensing is non-zero. But the fact that core images have not yet been
found (with perhaps one or two exceptions) is probably not a surprise.
As the search continues, two-image lenses should be better targets
than four-image lenses for revealing core images.

\acknowledgements
I am very grateful to Marijn Franx for discussions that contributed
to the birth and development of this project.
I also thank Martin Norbury for interesting discussions and for
providing data in advance of publication.
This work was supported by NASA through Hubble Fellowship grant
HST-HF-01141.01-A from the Space Telescope Science Institute, which
is operated by the Association of Universities for Research in
Astronomy, Inc., under NASA contract NAS5-26555.

\clearpage

\onecolumn

\begin{deluxetable}{lrrrrrrrrrrrrc}
\tabletypesize{\scriptsize}
\tablewidth{0pt}
\tablecaption{Galaxy Sample}
\tablehead{
  \colhead{(1)}
 &\colhead{(2)}
 &\colhead{(3)}
 &\colhead{(4)}
 &\colhead{(5)}
 &\colhead{(6)}
 &\colhead{(7)}
 &\colhead{(8)}
 &\colhead{(9)}
 &\colhead{(10)}
 &\colhead{(11)}
 &\colhead{(12)}
 &\colhead{(13)}
 &\colhead{(14)} \\
  \colhead{Name}
 &\colhead{$D$}
 &\colhead{$\sigma$}
 &\colhead{$\alpha$}
 &\colhead{$\beta$}
 &\colhead{$\gamma$}
 &\colhead{$\log\kappa_b$}
 &\colhead{$\log r_b$}
 &\colhead{$\log R_{\rm ein}$}
 &\colhead{$\log R_{\rm rad}$}
 &\colhead{$\log\ma$}
 &\colhead{$\log\ma$}
 &\colhead{$\log\ma$}
 &\colhead{Ref} \\
  \colhead{}
 &\colhead{(Mpc)}
 &\colhead{(km/s)}
 &\colhead{}
 &\colhead{}
 &\colhead{}
 &\colhead{}
 &\colhead{(arcsec)}
 &\colhead{(arcsec)}
 &\colhead{(arcsec)}
 &\colhead{}
 &\colhead{}
 &\colhead{}
 &\colhead{}
}
\startdata
NGC221  & $  0.9$ & $ 85$ & $ 0.98$ & $1.36$ & $0.01$ & $ 2.86$ & $-4.02$ & $-1.26$ & $-3.17$ & $-4.67$ & $-6.00$ & $-6.00$ & 1 \\
        & $  0.7$ & $ 77$ & $ 4.66$ & $1.26$ & $0.50$ & $ 2.56$ & $-3.71$ & $-1.29$ & $-3.28$ & $-4.74$ & $-6.00$ & $-5.31$ & 2 \\
NGC224  & $  0.9$ & $220$ & $ 4.72$ & $0.87$ & $0.12$ & $ 3.25$ & $-3.65$ & $ 0.43$ & $-0.58$ & $-1.65$ & $-1.66$ & $-1.66$ & 1 \\
NGC474  & $ 32.5$ & $169$ & $ 1.23$ & $1.90$ & $0.37$ & $ 0.91$ & $-1.39$ & $-0.43$ & $-1.30$ & $-2.08$ & $-2.10$ & $-2.10$ & 2 \\
NGC524  & $ 24.6$ & $275$ & $ 1.29$ & $1.00$ & $0.00$ & $ 1.96$ & $-2.21$ & $ 0.29$ & $-0.85$ & $-2.13$ & $-2.18$ & $-2.21$ & 1 \\
        & $ 32.1$ & $242$ & $ 0.68$ & $1.69$ & $0.03$ & $ 1.26$ & $-1.41$ & $-0.02$ & $-0.98$ & $-2.12$ & $-2.15$ & $-2.16$ & 2 \\
NGC596  & $ 22.6$ & $165$ & $ 0.76$ & $1.97$ & $0.55$ & $ 0.62$ & $-1.20$ & $-0.43$ & $-1.44$ & $-2.31$ & $-2.37$ & $-2.36$ & 1 \\
NGC720  & $ 24.1$ & $250$ & $ 2.32$ & $1.66$ & $0.06$ & $ 1.12$ & $-1.21$ & $-0.07$ & $-0.90$ & $-1.97$ & $-1.99$ & $-1.99$ & 1 \\
NGC821  & $ 23.2$ & $207$ & $ 1.00$ & $1.59$ & $0.64$ & $ 1.31$ & $-1.66$ & $-0.34$ & $-1.60$ & $-3.03$ & $-3.20$ & $-3.23$ & 2 \\
NGC1023 & $ 10.9$ & $217$ & $ 4.72$ & $1.18$ & $0.78$ & $ 1.52$ & $-1.80$ & $-0.16$ & $-1.52$ & $-2.99$ & $-3.11$ & $-3.14$ & 1 \\
NGC1052 & $ 17.8$ & $222$ & $ 1.05$ & $1.43$ & $0.11$ & $ 1.94$ & $-2.22$ & $-0.26$ & $-1.58$ & $-3.15$ & $-3.28$ & $-3.32$ & 2 \\
NGC1172 & $ 31.8$ & $113$ & $ 1.52$ & $1.64$ & $1.01$ & $ 0.18$ & $-1.21$ & $-0.76$ & $-6.00$ & $-6.00$ & $-6.00$ & $-6.00$ & 1 \\
NGC1316 & $ 19.1$ & $240$ & $ 1.16$ & $1.00$ & $0.00$ & $ 1.85$ & $-2.21$ & $ 0.20$ & $-0.87$ & $-1.96$ & $-1.99$ & $-2.01$ & 1 \\
NGC1399 & $ 19.1$ & $333$ & $ 1.50$ & $1.68$ & $0.07$ & $ 1.49$ & $-1.33$ & $ 0.10$ & $-0.95$ & $-2.66$ & $-2.69$ & $-2.72$ & 1 \\
NGC1400 & $ 22.9$ & $265$ & $ 1.39$ & $1.32$ & $0.00$ & $ 2.08$ & $-2.22$ & $-0.09$ & $-1.49$ & $-3.29$ & $-3.44$ & $-3.52$ & 1 \\
NGC1426 & $ 22.9$ & $150$ & $ 3.62$ & $1.35$ & $0.85$ & $ 0.89$ & $-1.53$ & $-0.51$ & $-1.52$ & $-2.25$ & $-2.31$ & $-2.30$ & 1 \\
NGC1600 & $ 53.5$ & $340$ & $ 1.98$ & $1.50$ & $0.08$ & $ 1.02$ & $-0.88$ & $ 0.26$ & $-0.52$ & $-1.68$ & $-1.69$ & $-1.70$ & 1 \\
NGC1700 & $ 37.9$ & $230$ & $ 0.90$ & $1.30$ & $0.00$ & $ 2.23$ & $-2.57$ & $-0.19$ & $-1.62$ & $-3.27$ & $-3.51$ & $-3.59$ & 1 \\
        & $ 54.1$ & $230$ & $ 0.46$ & $1.65$ & $0.01$ & $ 1.80$ & $-2.08$ & $-0.17$ & $-1.40$ & $-2.80$ & $-2.93$ & $-2.96$ & 3V \\
        & $ 54.1$ & $230$ & $ 0.47$ & $1.68$ & $0.01$ & $ 1.81$ & $-2.08$ & $-0.19$ & $-1.44$ & $-2.89$ & $-3.03$ & $-3.07$ & 3I \\
NGC2636 & $ 35.7$ & $ 85$ & $ 1.84$ & $1.14$ & $0.04$ & $ 1.41$ & $-2.59$ & $-0.88$ & $-1.87$ & $-1.97$ & $-2.01$ & $-1.99$ & 1 \\
NGC2685 & $ 16.2$ & $114$ & $ 1.69$ & $1.52$ & $0.73$ & $ 0.64$ & $-1.48$ & $-0.68$ & $-1.56$ & $-1.87$ & $-1.90$ & $-1.89$ & 2 \\
NGC2832 & $ 96.2$ & $330$ & $ 1.84$ & $1.40$ & $0.02$ & $ 1.27$ & $-1.16$ & $ 0.23$ & $-0.68$ & $-2.02$ & $-2.04$ & $-2.06$ & 1 \\
NGC2841 & $ 14.1$ & $229$ & $ 0.93$ & $1.02$ & $0.01$ & $ 2.35$ & $-2.84$ & $ 0.08$ & $-1.17$ & $-2.41$ & $-2.53$ & $-2.56$ & 1 \\
NGC3115 & $  9.0$ & $280$ & $ 1.47$ & $1.43$ & $0.78$ & $ 1.60$ & $-1.69$ & $-0.13$ & $-1.61$ & $-3.59$ & $-4.18$ & $-6.00$ & 1 \\
        & $  6.7$ & $264$ & $ 1.13$ & $1.80$ & $0.73$ & $ 1.36$ & $-1.51$ & $-0.25$ & $-1.65$ & $-3.62$ & $-4.00$ & $-6.00$ & 2 \\
NGC3377 & $ 10.6$ & $152$ & $ 1.92$ & $1.33$ & $0.29$ & $ 2.53$ & $-3.12$ & $-0.74$ & $-2.57$ & $-4.48$ & $-6.00$ & $-6.00$ & 1 \\
NGC3379 & $ 10.6$ & $225$ & $ 1.59$ & $1.43$ & $0.18$ & $ 1.61$ & $-1.84$ & $-0.21$ & $-1.35$ & $-2.72$ & $-2.78$ & $-2.80$ & 1 \\
        & $  8.1$ & $209$ & $ 1.82$ & $1.45$ & $0.18$ & $ 1.71$ & $-1.96$ & $-0.29$ & $-1.50$ & $-2.95$ & $-3.02$ & $-3.03$ & 2 \\
NGC3384 & $ 8.1$ & $170$ & $ 5.36$ & $1.58$ & $0.64$ & $ 1.35$ & $-1.78$ & $-0.52$ & $-1.61$ & $-2.80$ & $-2.84$ & $-2.84$ & 2 \\
NGC3599 & $ 21.7$ & $ 80$ & $13.01$ & $1.66$ & $0.79$ & $ 0.50$ & $-1.64$ & $-1.01$ & $-1.65$ & $-1.34$ & $-1.35$ & $-1.34$ & 1 \\
NGC3605 & $ 21.7$ & $103$ & $ 9.14$ & $1.26$ & $0.67$ & $ 0.92$ & $-1.82$ & $-0.75$ & $-1.64$ & $-1.84$ & $-1.86$ & $-1.85$ & 1 \\
NGC3608 & $ 21.7$ & $195$ & $ 1.05$ & $1.33$ & $0.00$ & $ 1.88$ & $-2.32$ & $-0.29$ & $-1.53$ & $-2.80$ & $-2.90$ & $-2.91$ & 1 \\
        & $ 13.6$ & $195$ & $ 0.72$ & $1.58$ & $0.00$ & $ 1.88$ & $-2.28$ & $-0.37$ & $-1.63$ & $-3.03$ & $-3.15$ & $-3.16$ & 3V \\
        & $ 13.6$ & $195$ & $ 0.78$ & $1.57$ & $0.00$ & $ 1.92$ & $-2.32$ & $-0.39$ & $-1.67$ & $-3.11$ & $-3.25$ & $-3.26$ & 3I \\
NGC3900 & $ 29.4$ & $118$ & $ 0.29$ & $1.66$ & $0.51$ & $ 1.39$ & $-2.24$ & $-0.75$ & $-2.37$ & $-3.65$ & $-6.00$ & $-6.00$ & 2 \\
NGC4026 & $ 17.0$ & $195$ & $ 0.88$ & $1.50$ & $0.68$ & $ 1.51$ & $-1.92$ & $-0.41$ & $-1.84$ & $-3.41$ & $-3.94$ & $-4.07$ & 2 \\
NGC4143 & $ 17.0$ & $270$ & $ 1.26$ & $2.18$ & $0.59$ & $ 1.20$ & $-1.35$ & $-0.26$ & $-1.45$ & $-3.23$ & $-3.35$ & $-3.42$ & 2 \\
NGC4150 & $  9.7$ & $ 85$ & $ 1.23$ & $1.67$ & $0.58$ & $ 1.18$ & $-2.28$ & $-1.09$ & $-2.20$ & $-2.69$ & $-2.79$ & $-2.74$ & 2 \\
NGC4168 & $ 38.8$ & $185$ & $ 0.95$ & $1.50$ & $0.14$ & $ 0.76$ & $-1.11$ & $-0.10$ & $-0.82$ & $-1.24$ & $-1.25$ & $-1.25$ & 1 \\
NGC4239 & $ 16.3$ & $ 60$ & $14.53$ & $0.96$ & $0.65$ & $ 0.39$ & $-1.78$ & $-1.09$ & $-1.71$ & $-0.75$ & $-0.76$ & $-0.75$ & 1 \\
NGC4261 & $ 35.1$ & $326$ & $ 2.38$ & $1.43$ & $0.00$ & $ 1.43$ & $-1.32$ & $ 0.15$ & $-0.86$ & $-2.38$ & $-2.41$ & $-2.44$ & 2 \\
NGC4278 & $  9.7$ & $250$ & $ 1.63$ & $1.39$ & $0.02$ & $ 1.96$ & $-2.10$ & $-0.16$ & $-1.50$ & $-3.25$ & $-3.36$ & $-3.40$ & 2 \\
        & $  9.7$ & $250$ & $ 1.45$ & $1.32$ & $0.00$ & $ 1.98$ & $-2.13$ & $-0.09$ & $-1.43$ & $-3.12$ & $-3.23$ & $-3.27$ & 3V \\
        & $  9.7$ & $250$ & $ 1.25$ & $1.46$ & $0.00$ & $ 1.90$ & $-2.04$ & $-0.15$ & $-1.43$ & $-3.13$ & $-3.22$ & $-3.25$ & 3I \\
NGC4291 & $ 29.4$ & $278$ & $ 2.07$ & $1.48$ & $0.02$ & $ 1.89$ & $-1.92$ & $-0.14$ & $-1.44$ & $-3.28$ & $-3.38$ & $-3.43$ & 2 \\
NGC4365 & $ 23.5$ & $262$ & $ 2.06$ & $1.27$ & $0.15$ & $ 1.43$ & $-1.51$ & $ 0.07$ & $-0.96$ & $-2.25$ & $-2.28$ & $-2.30$ & 1 \\
        & $ 13.8$ & $262$ & $ 1.67$ & $1.46$ & $0.11$ & $ 1.58$ & $-1.63$ & $-0.04$ & $-1.15$ & $-2.66$ & $-2.71$ & $-2.73$ & 3V \\
        & $ 13.8$ & $262$ & $ 1.52$ & $1.49$ & $0.04$ & $ 1.61$ & $-1.67$ & $-0.06$ & $-1.17$ & $-2.69$ & $-2.73$ & $-2.75$ & 3I \\
NGC4374 & $ 16.8$ & $296$ & $ 2.15$ & $1.50$ & $0.13$ & $ 1.53$ & $-1.47$ & $ 0.03$ & $-1.05$ & $-2.66$ & $-2.70$ & $-2.72$ & 2 \\
NGC4387 & $ 16.3$ & $105$ & $ 3.36$ & $1.59$ & $0.72$ & $ 0.38$ & $-1.24$ & $-0.66$ & $-1.33$ & $-1.24$ & $-1.26$ & $-1.25$ & 1 \\
NGC4406 & $ 16.8$ & $250$ & $ 3.31$ & $1.16$ & $0.00$ & $ 1.70$ & $-1.84$ & $ 0.03$ & $-1.19$ & $-2.61$ & $-2.67$ & $-2.69$ & 2 \\
        & $ 13.8$ & $250$ & $ 4.13$ & $1.05$ & $0.04$ & $ 1.80$ & $-1.95$ & $ 0.13$ & $-1.12$ & $-2.51$ & $-2.58$ & $-2.61$ & 3V \\
        & $ 13.8$ & $250$ & $ 3.32$ & $1.07$ & $0.00$ & $ 1.81$ & $-1.97$ & $ 0.12$ & $-1.14$ & $-2.55$ & $-2.63$ & $-2.66$ & 3I \\
NGC4417 & $ 16.8$ & $ 84$ & $ 0.87$ & $1.77$ & $0.71$ & $ 0.10$ & $-1.28$ & $-0.90$ & $-1.80$ & $-1.75$ & $-1.80$ & $-1.78$ & 2 \\
NGC4434 & $ 16.3$ & $115$ & $ 0.98$ & $1.78$ & $0.70$ & $ 0.60$ & $-1.51$ & $-0.74$ & $-1.77$ & $-2.31$ & $-2.39$ & $-2.35$ & 1 \\
NGC4458 & $ 16.3$ & $105$ & $ 5.26$ & $1.43$ & $0.49$ & $ 2.02$ & $-2.81$ & $-0.99$ & $-2.51$ & $-3.87$ & $-4.04$ & $-3.96$ & 1 \\
NGC4464 & $ 16.3$ & $125$ & $ 1.64$ & $1.68$ & $0.88$ & $ 0.96$ & $-1.81$ & $-0.81$ & $-2.08$ & $-3.11$ & $-6.00$ & $-3.35$ & 1 \\
NGC4467 & $ 16.3$ & $ 72$ & $ 7.52$ & $2.13$ & $0.98$ & $ 0.02$ & $-1.38$ & $-1.10$ & $-2.75$ & $-3.38$ & $-6.00$ & $-6.00$ & 1 \\
NGC4472 & $ 16.3$ & $300$ & $ 2.08$ & $1.17$ & $0.04$ & $ 1.51$ & $-1.51$ & $ 0.24$ & $-0.82$ & $-2.21$ & $-2.24$ & $-2.26$ & 1 \\
        & $ 16.8$ & $303$ & $ 1.89$ & $1.29$ & $0.04$ & $ 1.48$ & $-1.42$ & $ 0.20$ & $-0.84$ & $-2.29$ & $-2.32$ & $-2.34$ & 2 \\
NGC4478 & $ 16.3$ & $135$ & $ 3.32$ & $0.84$ & $0.43$ & $ 1.75$ & $-2.66$ & $-0.24$ & $-1.17$ & $-1.39$ & $-1.42$ & $-1.42$ & 1 \\
NGC4486 & $ 16.3$ & $360$ & $ 2.82$ & $1.39$ & $0.25$ & $ 1.32$ & $-1.01$ & $ 0.36$ & $-0.63$ & $-2.28$ & $-2.29$ & $-2.31$ & 1 \\
NGC4486B & $ 16.3$ & $200$ & $ 2.78$ & $1.33$ & $0.14$ & $ 2.24$ & $-2.63$ & $-0.50$ & $-2.10$ & $-3.90$ & $-4.33$ & $-4.38$ & 1 \\
NGC4551 & $ 16.3$ & $121$ & $ 2.94$ & $1.23$ & $0.80$ & $ 0.54$ & $-1.30$ & $-0.52$ & $-1.36$ & $-1.56$ & $-1.59$ & $-1.58$ & 1 \\
NGC4552 & $ 16.3$ & $260$ & $ 1.48$ & $1.30$ & $0.00$ & $ 1.93$ & $-2.08$ & $-0.05$ & $-1.36$ & $-3.02$ & $-3.12$ & $-3.16$ & 1 \\
        & $ 13.8$ & $260$ & $ 2.17$ & $1.06$ & $0.00$ & $ 2.06$ & $-2.25$ & $ 0.14$ & $-1.23$ & $-2.79$ & $-2.92$ & $-2.98$ & 3V \\
        & $ 13.8$ & $260$ & $ 2.10$ & $1.08$ & $0.04$ & $ 2.06$ & $-2.23$ & $ 0.12$ & $-1.27$ & $-2.87$ & $-3.00$ & $-3.06$ & 3I \\
NGC4564 & $ 16.3$ & $165$ & $ 0.25$ & $1.90$ & $0.05$ & $ 1.61$ & $-2.17$ & $-0.45$ & $-1.80$ & $-3.02$ & $-3.33$ & $-3.29$ & 1 \\
NGC4570 & $ 16.3$ & $195$ & $ 3.72$ & $1.49$ & $0.85$ & $ 1.04$ & $-1.44$ & $-0.36$ & $-1.43$ & $-2.57$ & $-2.65$ & $-2.65$ & 1 \\
NGC4589 & $ 30.0$ & $228$ & $ 1.09$ & $1.18$ & $0.11$ & $ 1.93$ & $-2.27$ & $-0.07$ & $-1.32$ & $-2.64$ & $-2.75$ & $-2.77$ & 2 \\
        & $ 24.4$ & $228$ & $ 0.43$ & $1.62$ & $0.00$ & $ 1.55$ & $-1.80$ & $-0.05$ & $-1.12$ & $-2.30$ & $-2.35$ & $-2.36$ & 3V \\
        & $ 24.4$ & $228$ & $ 0.50$ & $1.58$ & $0.01$ & $ 1.68$ & $-1.95$ & $-0.11$ & $-1.24$ & $-2.50$ & $-2.58$ & $-2.59$ & 3I \\
NGC4621 & $ 16.3$ & $250$ & $ 0.19$ & $1.71$ & $0.50$ & $ 1.16$ & $-1.42$ & $-0.10$ & $-1.90$ & $-3.94$ & $-6.00$ & $-6.00$ & 1 \\
NGC4636 & $ 16.3$ & $210$ & $ 1.64$ & $1.33$ & $0.13$ & $ 1.14$ & $-1.38$ & $-0.04$ & $-0.90$ & $-1.77$ & $-1.78$ & $-1.78$ & 1 \\
        & $ 17.0$ & $207$ & $ 1.69$ & $1.56$ & $0.13$ & $ 1.07$ & $-1.30$ & $-0.14$ & $-0.96$ & $-1.84$ & $-1.85$ & $-1.85$ & 2 \\
NGC4649 & $ 16.3$ & $360$ & $ 2.00$ & $1.30$ & $0.15$ & $ 1.55$ & $-1.34$ & $ 0.32$ & $-0.79$ & $-2.50$ & $-2.54$ & $-2.57$ & 1 \\
NGC4697 & $ 11.2$ & $175$ & $24.86$ & $1.04$ & $0.74$ & $ 1.27$ & $-1.64$ & $-0.11$ & $-1.26$ & $-2.24$ & $-2.28$ & $-2.28$ & 1 \\
NGC4742 & $ 13.3$ & $105$ & $48.60$ & $1.99$ & $1.09$ & $ 0.78$ & $-1.83$ & $-1.05$ & $-6.00$ & $-6.00$ & $-6.00$ & $-6.00$ & 1 \\
NGC4874 & $ 99.5$ & $290$ & $ 2.33$ & $1.37$ & $0.13$ & $ 0.71$ & $-0.68$ & $ 0.25$ & $-0.39$ & $-1.08$ & $-1.08$ & $-1.09$ & 1 \\
NGC4889 & $ 99.5$ & $350$ & $ 2.61$ & $1.35$ & $0.05$ & $ 1.06$ & $-0.88$ & $ 0.33$ & $-0.47$ & $-1.66$ & $-1.68$ & $-1.69$ & 1 \\
NGC5273 & $ 21.3$ & $ 52$ & $ 7.03$ & $1.32$ & $0.37$ & $ 0.48$ & $-1.93$ & $-1.26$ & $-1.79$ & $-0.72$ & $-0.73$ & $-0.73$ & 2 \\
NGC5813 & $ 30.2$ & $225$ & $ 2.15$ & $1.33$ & $0.08$ & $ 1.50$ & $-1.72$ & $-0.13$ & $-1.19$ & $-2.42$ & $-2.46$ & $-2.47$ & 1 \\
        & $ 21.2$ & $225$ & $ 1.77$ & $1.41$ & $0.03$ & $ 1.57$ & $-1.79$ & $-0.17$ & $-1.27$ & $-2.58$ & $-2.62$ & $-2.63$ & 3V \\
        & $ 21.2$ & $225$ & $ 1.67$ & $1.46$ & $0.01$ & $ 1.58$ & $-1.80$ & $-0.20$ & $-1.29$ & $-2.62$ & $-2.67$ & $-2.68$ & 3I \\
NGC5838 & $ 28.5$ & $290$ & $ 2.57$ & $1.87$ & $0.93$ & $ 0.83$ & $-0.98$ & $-0.12$ & $-1.29$ & $-2.97$ & $-6.00$ & $-6.00$ & 2 \\
NGC5845 & $ 30.1$ & $260$ & $ 1.27$ & $2.74$ & $0.51$ & $ 1.03$ & $-1.27$ & $-0.35$ & $-1.44$ & $-3.14$ & $-3.23$ & $-3.27$ & 1 \\
NGC5982 & $ 38.7$ & $256$ & $ 1.73$ & $1.28$ & $0.06$ & $ 1.65$ & $-1.80$ & $-0.02$ & $-1.16$ & $-2.57$ & $-2.62$ & $-2.65$ & 2 \\
        & $ 39.9$ & $256$ & $ 2.15$ & $1.19$ & $0.12$ & $ 1.64$ & $-1.79$ & $ 0.03$ & $-1.13$ & $-2.50$ & $-2.55$ & $-2.58$ & 3V \\
        & $ 39.9$ & $256$ & $ 2.17$ & $1.19$ & $0.12$ & $ 1.64$ & $-1.80$ & $ 0.03$ & $-1.13$ & $-2.51$ & $-2.57$ & $-2.59$ & 3I \\
NGC6166 & $120.0$ & $300$ & $ 3.32$ & $0.99$ & $0.08$ & $ 0.67$ & $-0.68$ & $ 0.35$ & $-0.27$ & $-0.74$ & $-0.75$ & $-0.76$ & 1 \\
NGC6340 & $ 22.0$ & $137$ & $ 2.46$ & $1.28$ & $0.59$ & $ 1.61$ & $-2.28$ & $-0.62$ & $-1.91$ & $-2.97$ & $-3.09$ & $-3.06$ & 2 \\
        & $ 18.0$ & $137$ & $ 1.73$ & $1.24$ & $0.72$ & $ 1.38$ & $-2.08$ & $-0.57$ & $-1.80$ & $-2.70$ & $-2.83$ & $-2.79$ & 4 \\
NGC7332 & $ 21.7$ & $130$ & $ 4.25$ & $1.34$ & $0.90$ & $ 1.12$ & $-1.88$ & $-0.68$ & $-1.87$ & $-2.76$ & $-2.93$ & $-2.86$ & 1 \\
NGC7457 & $ 12.3$ & $ 77$ & $ 2.32$ & $1.03$ & $0.35$ & $ 1.22$ & $-2.46$ & $-0.90$ & $-1.83$ & $-1.67$ & $-1.71$ & $-1.69$ & 2 \\
NGC7626 & $ 45.3$ & $273$ & $ 1.84$ & $1.30$ & $0.36$ & $ 1.57$ & $-1.64$ & $ 0.01$ & $-1.15$ & $-2.65$ & $-2.71$ & $-2.75$ & 2 \\
        & $ 49.5$ & $273$ & $ 1.53$ & $1.23$ & $0.00$ & $ 1.64$ & $-1.77$ & $ 0.08$ & $-1.03$ & $-2.40$ & $-2.45$ & $-2.47$ & 3V \\
        & $ 49.5$ & $273$ & $ 1.29$ & $1.27$ & $0.00$ & $ 1.65$ & $-1.78$ & $ 0.07$ & $-1.04$ & $-2.42$ & $-2.47$ & $-2.49$ & 3I \\
NGC7743 & $ 24.4$ & $ 85$ & $ 5.36$ & $1.38$ & $0.50$ & $ 1.47$ & $-2.48$ & $-1.04$ & $-2.19$ & $-2.76$ & $-2.82$ & $-2.79$ & 2 \\
NGC7768 & $110.0$ & $290$ & $ 1.92$ & $1.21$ & $0.00$ & $ 1.39$ & $-1.46$ & $ 0.17$ & $-0.82$ & $-2.04$ & $-2.07$ & $-2.08$ & 1 \\
\enddata
\tablecomments{
Col.~2: Distance (assuming $H_0 = 75$ km s$^{-1}$ Mpc$^{-1}$).
Col.~3: Central stellar velocity dispersion.
Cols.~4-6: Fitted Nuker law parameters.
Cols.~7-10: Logarithms of the lensing strength $\kappa_b$,
  break radius $r_b$, Einstein radius $R_{\rm ein}$, and radial critical
  radius $R_{\rm rad}$.
  Note that if any quantity is zero, we arbitrarily set its logarithm to
  $-6$.
Col.~11: Logarithm of mean core image magnification for the Nuker law lens.
Cols.~12-13: Logarithm of mean core image magnification for the Nuker law
  lens plus a supermassive black hole normalized by the $M_{\bullet}$--$\sigma$
  correlations of Gebhardt et al.\ (2000) and Merritt \& Ferrarese (2001),
  respectively.
Col.~14: References as follows:
(1) Faber et al.\ (1997);
(2) Ravindranath et al.\ (2001);
(3V) and (3I) V and I band samples, respectively, from Carollo et al.\ (1997);
(4) Carollo \& Stiavelli (1998).
}
\label{tab:sample}
\end{deluxetable}


\begin{references}

\reference{}
Barnes, J. E., \& Hernquist, L. 1992, \araa, 30, 705

\reference{}
Blandford, R. D., \& Kochanek, C. S. 1987, \apj, 321, 658

\reference{}
Blumenthal, G. R., Faber, S. M., Flores, R., \& Primack, J. R. 1986,
\apj, 3301, 27

\reference{}
Burke, W. L. 1981, \apj, 244, L1

\reference{}
Byun, Y.-I., et al. 1996, \aj, 111, 1889

\reference{}
Carollo, C. M., Franx, M., Illingworth, G. D., \& Forbes, D. A. 1997,
\apj, 481, 710

\reference{}
Carollo, C. M., \& Stiavelli, M. 1998, \aj, 115, 2306

\reference{}
Chen, G. H., \& Hewitt, J. N. 1993, \aj, 106, 1719

\reference{}
de Zeeuw, T., \& Franx, M. 1991, \araa, 29, 239

\reference{}
Ebisuzaki, T., Makino, J., \& Okumura, S. K. 1991, \nat, 354, 212

\reference{} 
Egami, E., Neugebauer, G., Soifer, B. T., Matthews, K., Ressler, M.,
Becklin, E. E., Murphy, T. W., \& Dale, D. A. 2000, \apj, 535, 561

\reference{}
Evans, N. W., \& Hunter, C. 2002, preprint (astro-ph/0204206)

\reference{}
Faber, S., et al. 1997, \aj, 114, 1771

\reference{}
Gebhardt, K., et al. 2000, \apj, 539, L13

\reference{}
Gerhard, O., Kronawitter, A., Saglia, R. P., \& Bender, R. 2001, \aj,
121, 1936

\reference{}
Hernquist, L. 1990, ApJ, 356, 359

\reference{}
Hinshaw, G., \& Krauss, L. M. 1987, \apj, 320, 468

\reference{} 
Ibata, R. A., Lewis, G. F., Irwin, M. J., Leh\'ar, J., \& Totten, E. J.
1999, \aj, 118, 1922

\reference{}
Kayser, R., Refsdal, S., \& Stabell, R. 1986, \aap, 166, 36

\reference{} 
Keeton, C. R. 2001, \apj, 561, 46

\reference{} 
Kochanek, C. S. 1996, \apj, 466, 638

\reference{} 
Kochanek, C. S, Falco, E. E., Impey, C. D., Leh\'ar, J., McLeod, B. A.,
Rix, H.-W., Keeton, C. R., Mu\~noz, J. A., \& Peng, C. Y. 2000, \apj,
543, 131

\reference{}
Lauer, T. R., et al. 1995, \aj, 110, 2622

\reference{} 
Lewis, G., Carilli, C., Papadopoulos, P., \& Ivison, R. J. 2002,
\mnras, 330, L15

\reference{}
Magorrian, J., et al. 1998, \aj, 115, 2285

\reference{}
Mao, S., Witt, H. J., \& Koopmans, L. V. E. 2001, \mnras, 323, 301

\reference{}
Mellier, Y., Fort, B., \& Kneib, J.-P. 1993, \apj, 407, 33

\reference{}
Merritt, D., \& Ferrarese, L. 2001, \apj, 547, 140

\reference{}
Milosavljevic, M., \& Merritt, D. 2001, \apj, 563, 34

\reference{}
Milosavljevic, M., Merritt, D., Rest, A., \& van den Bosch, F. C. 2001,
preprint (astro-ph/0110185)

\reference{}
Molikawa, K., \& Hattori, M. 2001, \apj, 559, 544

\reference{}
Mu\~noz, J. A., Kochanek, C. S., \& Keeton, C. R. 2001, \apj, 558, 657

\reference{}
Narasimha, D., Subramanian, K., \& Chitre, S. M. 1986, \nat, 321, 45

\reference{}
Narayan, R., Blandford, R., \& Nityananda, R. 1984, \nat, 310, 112

\reference{}
Narayan, R., \& Schneider, P. 1990, \mnras, 243, 192

\reference{}
Norbury, M. A., Rusin, D., Jackson, N., Browne, I. W. A., \&
Wilkinson, P. N. 2002, \mnras, submitted

\reference{}
Oguri, M., Taruya, A., \& Suto, Y. 2001, \apj, 559, 572

\reference{}
Ravindranath, S., Ho, L. C., Peng, C. Y., Filippenko, A. V., \&
Sargent, W. L. W. 2001, \aj, 122, 653

\reference{}
Rest, A., van den Bosch, F. C., Jaffe, W., Tran, H., Tsvetanov, Z.,
Ford, H. C., Davies, J., \& Schafer, J. 2001, \aj, 121, 2431

\reference{}
Rusin, D., et al. 2001, \apj, 557, 594

\reference{}
Rusin, D., \& Ma, C.-P. 2001, \apj, 549, L33

\reference{}
Schneider, P., Ehlers, J., \& Falco, E. E. 1992, Gravitational Lenses
(New York: Springer)

\reference{}
Schramm, T. 1990, \aap, 231, 19

\reference{}
Smail, I., Dressler, A., Kneib, J.-P., Ellis, R. S., Couch, W. J.,
Sharples, R. M., \& Oemler, A. 1996, \apj, 469, 508

\reference{}
Spergel, D. N., \& Steinhardt, P. J. 2000, \prl, 84, 3760

\reference{}
Tremaine, S. 1997, in Unsolved Problems in Astrophysics, ed. J. N. Bahcall
\& J. P. Ostriker (Princeton: Princeton Univ. Press), 137

\reference{}
Treu, T., \& Koopmans, L. V. E. 2002, preprint (astro-ph/0202342)

\reference{}
Wallington, S., \& Narayan, R. 1993, \apj, 403, 517

\reference{}
Wambsganss, J. 1997, \mnras, 284, 172

\reference{}
Winn, J. N., Morgan, N. D., Hewitt, J. N., Kochanek, C. S., Lovell, J. E. J.,
Patnaik, A. R., Pindor, B., Schechter, P. L., \& Schommer, R. A. 2002,
\aj, 123, 10

\end{references}
\end{document}